\documentclass[letterpaper,twocolumn,10pt]{article}
\usepackage{usenix-2020-09}

\usepackage{tikz}
\usepackage{amsmath}

\makeatletter
\usepackage{color, pifont, comment}
\usepackage{xspace}
\usepackage{xcolor}
\usepackage[ligature, inference]{semantic}
\usepackage{amsmath}
\usepackage{graphicx}
\usepackage{listings}
\usepackage{color, colortbl}
\usepackage{textcomp}
\usepackage{multirow}
\usepackage{makecell}
\usepackage{hyperref}
\usepackage[ruled, vlined, linesnumbered]{algorithm2e}
\usepackage[all]{xy}
\usepackage{tabularx}
\usepackage{caption}
\usepackage{subcaption}
\usepackage{booktabs}
\usepackage{tikz}

\newcounter{isusenixversion}
\ifthenelse{\value{isusenixversion}>0}
{
    \usepackage[available,functional,reproduced]{usenixbadges}
}

\newcommand{\ie}{{\it i.e.,}\xspace}
\newcommand{\eg}{{\it e.g.,}\xspace}
\newcommand{\etc}{{\it etc.}\xspace}

\newcommand{\etal}{{\it et al.}\xspace}
\newcommand{\vs}{{\it vs.}\xspace}
\newcommand{\wrt}{{\it w.r.t.}\xspace}
\newcommand{\descr}[1]{\vspace{.05in} \noindent\textbf{#1}}
\newcommand{\tool}{{\scshape OVRseen}}

\newcommand{\netdataset}{network traffic\xspace dataset}
\newcommand{\device}{{Quest 2}}

\newcommand{\topoculusfree}{Oculus-Free}
\newcommand{\topoculuspaid}{Oculus-Paid}
\newcommand{\topsidequest}{SideQuest}

\newcommand{\esld}{eSLD}
\newcommand{\tshark}{\code{tshark}}

\newcommand{\blocklist}{blocklist}
\newcommand{\policheckdataflow}{\textit{$\langle$data type, entity$\rangle$}}
\newcommand{\dataflow}{\textit{$\langle$app, data type, destination$\rangle$}}
\newcommand{\dataflowtuple}{\textit{$\langle$app, data type, destination$\rangle$}}
\newcommand{\polichecktuple}{\textit{$\langle$app, data type, entity$\rangle$}}
\newcommand{\dataflowtotal}{1,135}
\newcommand{\packettotal}{7,775}
\newcommand{\tcpflowtotal}{6,269}
\newcommand{\appstotal}{140}

\newcommand{\policheck}{PoliCheck}
\newcommand{\polisys}{Polisis}
\newcommand{\policylint}{PolicyLint}

\newcommand{\secref}{Section}
\newcommand{\figref}{Fig.}
\newcommand{\tabref}{Table}
\newcommand{\appref}{Appendix}

\newcommand{\oculusvr}{OVR}

\definecolor{maroon}{cmyk}{0,0.87,0.68,0.32}
\definecolor{brandeisblue}{rgb}{0.0, 0.44, 1.0}

\newcommand{\tabletextsize}{\footnotesize}

\newcommand{\code}[1]{\text{\tt #1}}

\newcommand{\todo}[1]{{\color{red} \bf [[#1]]}}
\newcommand{\mysection}[1]{\section{#1}}

\newcommand{\myparagraph}[1]{\vspace{-10pt}\paragraph{#1}}

\newcommand{\mycomment}[1]{}

\newcommand*\circled[1]{\tikz[baseline=(char.base)]{
            \node[shape=circle,draw,inner sep=1pt] (char) {\scriptsize#1};}}

\newcommand{\squishcount}{
   \begin{list}{\arabic{enumi})}
     { \usecounter{enumi}
       \setlength{\itemindent}{0em}
       \setlength{\parskip}{0pt}
      \setlength{\itemsep}{0pt}      \setlength{\parsep}{1pt}
      \setlength{\topsep}{1pt}       \setlength{\partopsep}{0pt}
      \setlength{\leftmargin}{1.5em} \setlength{\labelwidth}{10em}
      \setlength{\labelsep}{0.5em} } }
\newcommand{\countend}{
  \end{list}}

\newcommand{\minorrevision}[1]{{\color{black} #1}}

\newcommand{\rahmadi}[1]{{\color{cyan} #1}}
\newcommand{\ashuba}[1]{{\color{green}#1}}
\newcommand{\cuihao}[1]{{\color{violet} #1}}
\newcommand{\hieu}[1]{{\color{orange}#1}}
\newcommand{\athina}[1]{{\color{red}#1}}
\newcommand{\term}[1]{{\color{magenta}#1}}

\ifthenelse{\value{isusenixversion}>0}
{
    \pagestyle{empty} 
    \newcommand{\appdx}[1]{\text{\color{black} #1 in~\cite{trimananda2021auditing}}}
    \newcommand{\inusenixversion}[1]{#1}
    \newcommand{\inarxivversion}[1]{}
}
{
    \newcommand{\appdx}[1]{#1}
    \newcommand{\inusenixversion}[1]{}
    \newcommand{\inarxivversion}[1]{#1}
}

\begin{document}

\date{}

 \title{\tool{}: Auditing Network Traffic and Privacy Policies in Oculus VR}

\mycomment{
\author{
{\rm Rahmadi Trimananda,* Hieu Le,* Hao Cui,* Janice Tran Ho,* Anastasia Shuba,\dag{} and Athina Markopoulou*}\\
\\
*University of California, Irvine\\
{\rm \dag{}}Independent Researcher\\
} 
}

\author{
{\rm Rahmadi Trimananda,\textsuperscript{1} Hieu Le,\textsuperscript{1} Hao Cui,\textsuperscript{1} Janice Tran Ho,\textsuperscript{1} Anastasia Shuba,\textsuperscript{2} and Athina Markopoulou\textsuperscript{1}}\\
\\
{\rm \textsuperscript{1}University of California, Irvine}\\
{\rm \textsuperscript{2}Independent Researcher}\\
} 

\maketitle

\begin{abstract}

Virtual reality (VR) is an emerging technology that enables new applications but also introduces privacy risks. In this paper, we focus on Oculus VR (\oculusvr{}), the leading platform in the VR space and we provide the first comprehensive analysis of personal data exposed by \oculusvr{} apps and the platform itself, 
from a combined networking and privacy policy perspective.  We experimented with the \device{} headset and tested the most popular VR apps available on the official Oculus and the Side\-Quest app stores.  We developed \tool{}, a methodology and system for collecting, analyzing, and comparing network traffic and privacy policies on \oculusvr{}. On the networking side, we captured and decrypted network traffic of VR apps, which was previously not possible on \oculusvr{}, and we extracted data flows, defined as \dataflowtuple{}. 
Compared  to the mobile and other app ecosystems, we found \oculusvr{} to be more centralized and driven by tracking and analytics, rather than by third-party advertising. 
We show that the data types exposed by VR apps include personally identifiable information (PII), device information that can be used for fingerprinting, and VR-specific data types. By comparing the data flows found in the network traffic with statements made in the apps' privacy policies, we
found that approximately 70\% of \oculusvr{} data flows were not properly disclosed. Furthermore, we extracted additional context 
from the privacy policies, and we observed that 69\% of the data flows were used for purposes unrelated to the core functionality of apps.  

\end{abstract}

\mycomment{
\begin{figure*}[ht!]
	\centering
	\includegraphics[width=0.8\textwidth]{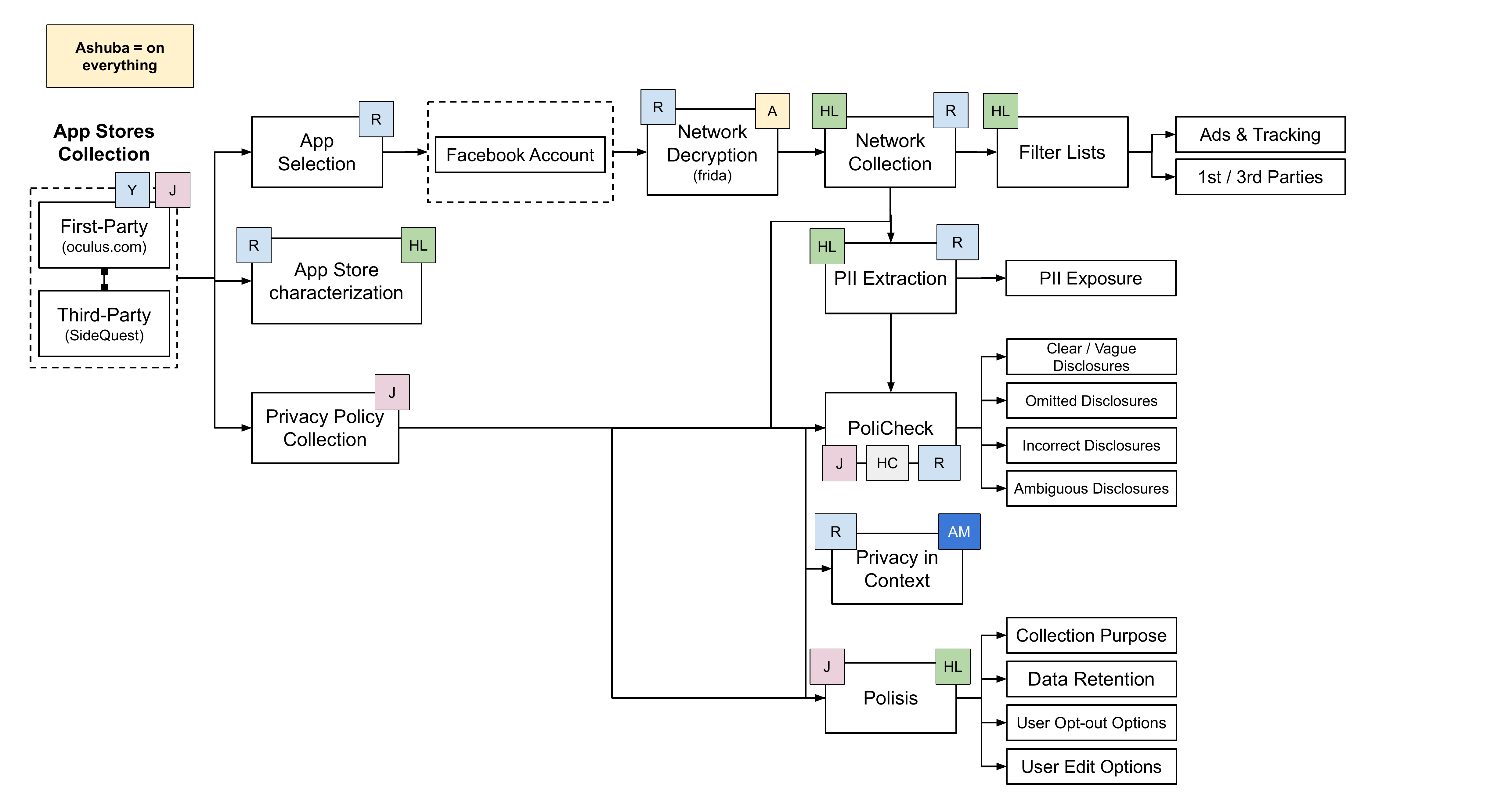}
	\caption{\tool{} Workflow: TBD}
	\label{fig:main-workflow}
\end{figure*}
}

\section{Introduction}
\label{introduction}
Virtual reality (VR) technology has created an emerging market: VR hardware and software revenues are projected to increase from \$800 million in 2018 to \$5.5 billion in 2023~\cite{vr-market}. 
%
%
Among VR platforms, Oculus VR (\oculusvr{}) is a pioneering, and arguably the most popular one: within six months since October 2020, an estimated five million \device{} headsets were sold~\cite{oculus-acquisition, oculus2-popular}.
\mycomment{
Oculus VR (\oculusvr{}), in particular, is a pioneering, and arguably the most popular, VR platform: within six months since October 2020, an estimated five million \device{} headsets were sold~\cite{oculus-acquisition,fb-q1-2021-earnings, oculus2-popular, oculus2-sales}.}
VR technology enables a number of applications, including recreational games, physical training, health therapy, and many others~\cite{vr-use-cases}.

VR also introduces privacy risks: some are similar to those on other Internet-based platforms (\eg mobile phones~\cite{androidappstudy,taintdroid}, IoT devices~\cite{iotsok,iotsecurity}, and Smart TVs~\cite{mohajeri2019watching,varmarken2020tv}), but others are unique to VR. %
%
For example, VR headsets and hand controllers 
are equipped with sensors that may collect data about the user's physical movement, body characteristics and activity, voice activity, hand tracking, eye tracking, facial expressions, and play area~\cite{oculus-privacy-1,oculus-privacy-3}, which may in turn reveal information about our physique, emotions, and home.
\mycomment{For example, VR headsets and hand controllers 
 are equipped with sensors that may collect data about the user's physical movement, body characteristics and activity, voice activity, hand tracking, eye tracking, facial expressions, and play area~\cite{oculus-privacy-1,oculus-privacy-2,oculus-privacy-3}, which may in turn reveal information about our physique, emotions, and home.}
The privacy aspects of VR platforms are currently not well understood~\cite{adams2018ethics}.
\begin{figure*}[t!]
	\centering
	\vspace{-5pt}
	\includegraphics[width=0.9\linewidth]{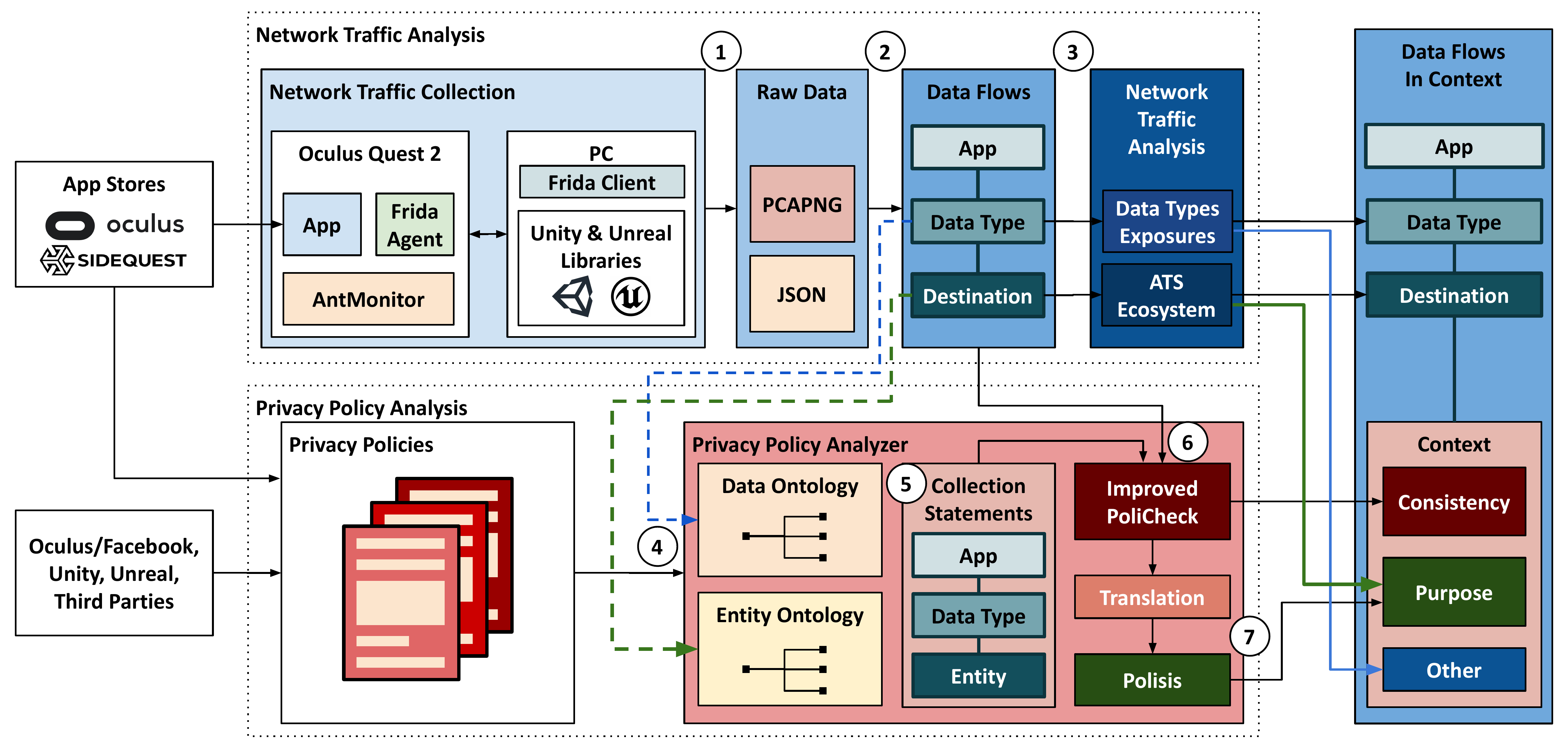}
	\vspace{-5pt}
	\caption{\textbf{Overview of \tool{}.}	We select the most popular apps from the official Oculus and SideQuest app stores. 
    First,	we experiment with them and analyze their {\bf network traffic:} (1) we obtain raw data in PCAPNG and JSON; (2) we extract data flows \dataflowtuple{}; and (3) we analyze them \wrt{} data types and ATS ecosystem. 
	Second, we analyze the same apps' (and their used libraries') {\bf privacy policies}: (4) we build VR-specific data and entity ontologies, informed both by network traffic and privacy policy text; and (5) we extract collection statements \polichecktuple{}  from the privacy policy. Third, we {\bf compare the two}: (6) using our improved \policheck{}, we map each data flow to a collection statement, and we perform network-to-policy consistency analysis. Finally,  (7) we  translate the sentence containing the collection statement into a text segment that \polisys{} can use to extract the data collection purpose. The end result is that data flows, extracted from network traffic, are augmented with additional {\bf context}, such as consistency with policy and purpose of collection. 
	}
	\label{fig:our-work}
	\vspace{-10pt}
\end{figure*}

To the best of our knowledge, our work is the first large scale, comprehensive measurement and characterization of privacy aspects of \oculusvr{} apps and platform, from a combined network and privacy policy point of view. We set out to characterize how sensitive information is collected and shared in the VR ecosystem, in theory (as described in the privacy policies) and in practice (as exhibited in the network traffic generated by VR apps). We center our analysis around the concept of {\em data flow}, which we define as the tuple \dataflowtuple~extracted from the network traffic. First, we are interested in the sender of information, namely the VR \emph{app}.  
Second, we are interested in the exposed {\em data types}, including personally identifiable information (PII), device information that can be used for fingerprinting, and VR sensor data. 
Third, we are interested in the recipient of the information, namely the {\em destination} domain, which we further categorize into entity or organization, first \vs{} third party \wrt{} the sending app, and ads and tracking services (ATS).  
%
Inspired by the framework of contextual integrity~\cite{privacy-in-context}, we also seek to characterize whether the data flows are appropriate or not within their context. More specifically, our notion of context includes: {\em consistency}, \ie whether actual data flows extracted from network traffic agree  with the corresponding statements made in the privacy policy; {\em purpose}, extracted from privacy policies and confirmed by destination domains (\eg whether they are ATS); and other information (\eg ``notice and consent'').  
Our methodology and system, \tool{}, 
is depicted on \figref{}~\ref{fig:our-work}. Next we summarize our methodology and findings.

\vspace{5pt}
\myparagraph{Network traffic: methodology and findings.} 
\minorrevision{We were able to explore 140 popular, paid and free, \oculusvr{} apps; and then capture, decrypt, and analyze the network traffic they generate in order to assess their practices with respect to collection and sharing of personal data on the \oculusvr{} platform.} 

%
\tool{} collects network traffic without rooting the \device{}, by building on the open-source  AntMonitor~\cite{shuba2016antmonitor}, which we had to modify to work on the Android 10-based Oculus OS. 
\minorrevision{
Furthermore, we successfully addressed new technical challenges for decrypting network traffic on \oculusvr{}. \tool{} combines  dynamic analysis (using} Frida~\cite{frida-tool}) \minorrevision{with binary analysis to find and bypass certificate validation functions, even when the app contains a stripped binary~\cite{Unity-strip}.} This was a  challenge specific to \oculusvr{}: prior work on decrypting network traffic on Android~\cite{shuba2020nomoats,mohajeri2019watching} hooked into standard Android SDK functions and not the ones packaged with Unity and Unreal, which are the basis for game apps.

We extracted and analyzed data flows found in the collected network traffic from the 140 OVR apps, and we made the following observations. 
We studied a broad range of 21 data types that are exposed and found that 33 and 70 apps send PII data types (\eg Device ID, User ID, and Android ID) to first-and third-party destinations, respectively (see Table~\ref{tab:pii-summary}). Notably, 58 apps expose VR sensory data (\eg physical movement, play area) to third-parties.
\minorrevision{
We used state-of-the-art \blocklist{s} to identify ATS and discovered that, unlike other popular platforms (\eg{} Android and Smart TVs), \oculusvr{} exposes data primarily to tracking and analytics services, and has a less diverse tracking ecosystem. Notably, the \blocklist{s} identified only 36\% of these exposures. On the other hand, we found no data exposure to advertising services as ads on \oculusvr{} is still in an experimental phase~\cite{oculus-ads}. 
}
%
%

\myparagraph{Privacy policy: methodology and findings.}
We provide an NLP-based methodology for analyzing the privacy policies that accompany VR apps. More specifically, \tool{}  maps each data flow (found in the network traffic) to its corresponding  data collection statement (found in the text of the privacy policy), and checks the {\em consistency} of the two. Furthermore, it extracts the {\em purpose} of data flows from the privacy policy, as well as from the ATS analysis of destination domains. Consistency, purpose, and additional information provide {\em context}, in which we can holistically assess the appropriateness of a data flow \cite{privacy-in-context}. Our methodology builds on, combines, and  improves state-of-the-art tools originally developed for mobile apps:  \policylint{}~\cite{andow2019policylint}, \policheck{}~\cite{andow2020actions}, and \polisys{}~\cite{harkous2018polisis}.
We curated VR-specific ontologies for data types and entities, guided by both the network traffic and privacy policies. We also interfaced between different NLP models of \policheck{} and \polisys{} to extract the purpose behind each data flow. 

%

%
Our network-to-policy consistency analysis revealed that about 70\% of data flows from  VR apps were not disclosed or consistent with their privacy policies\minorrevision{: only 30\% were consistent}. Furthermore, 38 apps did not have privacy policies, including apps from the official Oculus app store. Some app developers also had the tendency to neglect declaring data collected by the platform and third parties.
\minorrevision{We found that by automatically including these other parties' privacy policies in \tool{}'s network-to-policy consistency analysis, 74\% of data flows became consistent.
}
We also found that 69\% of data flows have purposes unrelated to the core functionality \minorrevision{(\eg advertising, marketing campaigns, analytics)}, and only a handful of apps are explicit about notice and consent. 
\minorrevision{\tool{}'s implementation and datasets are made available at~\cite{ovrseen-release}.} 
\myparagraph{Overview.} The rest of this paper is structured as follows.
\secref{}~\ref{sec:background} provides background on the \oculusvr{} platform and its data collection practices that motivate our work. 
\secref{}~\ref{sec:network-traffic-analysis} provides the methodology and results for \tool{}'s network traffic  analysis. 
\secref{}~\ref{sec:privacy-policy} provides the methodology and results for \tool{}'s policy analysis, network-to-policy consistency analysis, and purpose extraction.
\minorrevision{\secref{}~\ref{sec:discussion} discusses 
the findings and provides recommendations.}
\secref{}~\ref{sec:related-work} discusses related work.
\secref{}~\ref{sec:limitation-conclusion} concludes the paper.

\section{ Oculus VR Platform and Apps}
\label{sec:background}

\mycomment{
In this section, we first provide the background on \oculusvr{} (\secref{}~\ref{sec:platform-motivation}). Further, we also discuss how data collection and privacy policy practices evolve on \oculusvr{}---this motivates our study on the privacy of VR platform (\secref{}~\ref{sec:oculusdataprivacy}).

\subsection{\oculusvr{} Platform}
\label{sec:platform-motivation}
}

In this paper, we focus on the Oculus VR (\oculusvr{}), a representative of state-of-the art VR platform. A pioneer and leader in the VR space, \oculusvr{} was bought by Facebook in 2014~\cite{oculus-acquisition} (we refer to both as ``platform-party''), and it maintains to be the most popular VR platform today.  
Facebook has integrated more social features and analytics to \oculusvr{} and now even requires users to sign in using a Facebook account~\cite{oculus-blog-force-login}.

We used the latest Oculus device, \device{}, for testing. 
\device{} is completely wireless: it can operate standalone and run apps, without being connected to other devices. In contrast, \eg Sony Playstation VR needs to be connected to a Playstation 4 as its game controller.
\device{} runs Oculus OS, a variant of Android 10 that has been modified and optimized to run VR environments and apps.
The device comes with a few pre-installed apps, such as the Oculus browser. 
VR apps are usually developed using two popular game engines called Unity~\cite{unity-framework} and Unreal~\cite{unreal-framework}. 
%
Unlike traditional Android apps that run on Android JVM, these 3D app development frameworks compile VR apps into optimized (\ie stripped) binaries to run on \device{}~\cite{Unity-strip}.

Oculus has an official app store and a number of third-party app stores.
The Oculus app store offers a wide range of apps (many of them are paid), which are carefully curated and tested (\eg for VR motion sickness). 
In addition to the Oculus app store, we focus on SideQuest---the most popular third-party app store endorsed by Facebook~\cite{oculus-sidequest}.
In contrast to apps from the official store, apps available on SideQuest are typically at their early development stage and thus are mostly free. 
Many of them transition from SideQuest to the Oculus app store once they mature and become paid apps.
%
%
As of March 2021, the official Oculus app store has 267 apps (79 free and 183 paid), and
the SideQuest app store has 1,075 apps (859 free and 218 paid).

\myparagraph{Motivation: privacy risks in \oculusvr{}.} VR introduces privacy risks, some of which are similar to other Internet-based platforms \minorrevision{(\eg Android}~\cite{androidappstudy,taintdroid}, IoT devices~\cite{iotsok,iotsecurity}, Smart TVs~\cite{mohajeri2019watching,varmarken2020tv}), \etc),   
while others are unique to the VR platform. For example, VR headsets and hand controllers 
 are equipped with sensors that collect data about the user's physical movement, body characteristics, voice activity, hand tracking, eye tracking, facial expressions, and play area~\cite{oculus-privacy-1,oculus-privacy-2,oculus-privacy-3}, which may in turn reveal sensitive  information about our physique, emotions, and home.  
\device{} can also act as a fitness tracker, thanks to the built-in Oculus Move app that tracks time spent for actively moving and amount of calories burned across all apps~\cite{oculus-move}.
\mycomment{\device{} can also act as a fitness tracker, thanks to the built-in Oculus Move app that tracks time spent for actively moving and amount of calories burned across all apps~\cite{forbes-fitness,oculus-move}.}
 Furthermore, Oculus has been continuously updating their privacy policy with a trend of increasingly collecting more data over the years. Most notably, we observed a major update in May 2018, coinciding with the GDPR implementation date.
Many apps have no privacy policy, or fail to properly include the privacy policies of third-party libraries. Please see \appdx{\appref{}~\ref{sec:oculusdataprivacy}} for more detail on observations that motivated our study, and \secref{}~\ref{sec:related-work} on related work. The privacy risks on the relatively new VR platform are not yet well understood. 


\myparagraph{Goal and approach: privacy analysis of \oculusvr.} In this paper, we seek to characterize the privacy risks introduced when potentially-sensitive data available on the device are sent by the VR apps and/or the platform to remote destinations for various purposes. We followed an experimental and data-driven approach, and we chose to test and analyze the most popular VR apps.
In \secref~\ref{sec:network-traffic-analysis}, we characterize the actual behavior exhibited in the network traffic generated by these VR apps and platform. In \secref~\ref{sec:privacy-policy}, we present how we downloaded the privacy policies of the selected VR apps, the platform, and relevant third-party libraries, used NLP to extract and analyze the statements made about data collection, analyzed their consistency when compared against the actual data flows found in traffic, and extracted the purpose of data collection.

\myparagraph{App corpus.}
We selected \oculusvr{} apps that are widely used by players. 
Our app corpus consists of \minorrevision{150} popular paid and free apps from both the official Oculus app store and SideQuest. In contrast,  previous work typically considered only free apps from the official app store~\cite{androidappstudy,taintdroid,mohajeri2019watching,varmarken2020tv}.
We used the number of ratings/reviews as the popularity metric, and considered only apps that received at least 3.5 stars. We selected \minorrevision{three groups of 50 apps each: (1) the top-50  free apps and (2) the top-50 paid apps from the Oculus app store, and (3) the top-50 apps from the SideQuest store.  We selected an equal number of paid and free apps from the Oculus app store to gain insight into both groups equally.} We purposely did not just pick the top-100 apps, because paid apps tend to receive more reviews from users and this would bias our findings towards paid apps. Specifically, this would make our corpus consist of 90\% paid and 10\% free apps.

Our app corpus is representative of both app stores.
Our top-50 free and top-50 paid Oculus apps constitute close to 40\% of all apps on the Oculus app store, whereas the total number of downloads of our top-50 SideQuest apps is approximately 45\% of all downloads for the SideQuest store.
%
\minorrevision{Out of these 150 apps, selected for their popularity and representativeness, we were able to decrypt and analyze the network traffic for 140 of them for reasons explained in Section \ref{sec:network-traffic-dataset-raw}.}

\section{\tool{}: Network Traffic}
\label{sec:network-traffic-analysis}

In this section, we detail our methodology for collecting and analyzing network traffic. 
\minorrevision{
In \secref~\ref{sec:network-traffic-collection}, we present \tool{}'s system for collecting network traffic and highlight our decryption technique.}
Next, in \secref~\ref{sec:network-traffic-dataset}, we describe our network traffic dataset and the extracted data flows.
%
In \secref{}~\ref{sec:ats-ecosystem}, we report our findings on the \oculusvr{} ATS ecosystem by identifying domains that were labeled as ATS by popular blocklists. 
%
Finally, in \secref{}~\ref{sec:pii-exposures}, we discuss data types exposures in the extracted data flows according to the context based on whether their destination is an ATS or not.

\subsection{Network Traffic Collection}
\label{sec:network-traffic-collection}
In this section, we present \tool{}'s system for collecting and decrypting the network traffic that apps generate (\circled{\textbf{1}} in \figref{}~\ref{fig:our-work}). 
It is important to mention that \tool{} does not require rooting \device{}, and as of June 2021, there are no known methods for doing so~\cite{failedjailbreak}.
Since the Oculus OS is based on Android, we enhanced AntMonitor~\cite{shuba2016antmonitor} to support the Oculus OS.
Furthermore, to decrypt TLS traffic, we use 
Frida~\cite{frida-tool}, a dynamic instrumentation toolkit. 
\minorrevision{
 Using Frida to bypass certificate validation specifically for \device{} apps presents new technical challenges, compared to Android apps that have a different structure.
Next, we describe these challenges and how we address them.
}

\myparagraph{Traffic collection.}
For collecting network traffic, \tool{} integrates AntMonitor~\cite{shuba2016antmonitor}---a VPN-based tool for Android that does not require root access. It runs completely on the device without the need to re-route traffic to a server.
AntMonitor stores the collected traffic in PCAPNG format, where each packet is annotated (in the form of a PCAPNG comment) with the name of the corresponding app. 
To decrypt TLS connections, AntMonitor installs a user CA certificate. 
However, since Oculus OS is a modified version of Android 10, and AntMonitor only supports up to Android 7, we made multiple compatibility changes to support Oculus OS.
In addition, we enhanced the way AntMonitor stores decrypted packets: we adjust the sequence and ack numbers to make packet re-assembly by common tools (\eg \tshark) feasible in post-processing.
We will submit a pull request to AntMonitor's open-source repository, so that other researchers can make use of it, not only on \device{}, but also on other newer Android devices. 
For further details, see \appdx{\appref{}~\ref{app:antmonitor-improvement}}.


\myparagraph{TLS decryption.}
Newer Android devices, such as \device{}, pose a challenge for TLS decryption: as of Android 7, apps that target API level 24 (Android 7.0) and above no longer trust user-added certificates \cite{android7tls}.
Since \device{} cannot be rooted, we cannot install AntMonitor's certificate as a system certificate.
Thus, to circumvent the mistrust of AntMonitor's certificate, \tool{} uses Frida (see \figref{}~\ref{fig:our-work}) to intercept certificate validation APIs. 
To use Frida in a non-rooted environment, we extract each app and repackage it to include and start the Frida server when the app loads. 
The Frida server then listens to commands from a Frida client that is running on a PC using ADB.
Although ADB typically requires a USB connection, we run ADB over TCP to be able to use \device{} wirelessly, allowing for free-roaming testing of VR apps.

\tool{} uses the Frida client to load and inject our custom JavaScript code that intercepts various APIs used to verify CA certificates.
In general, Android and \device{} apps use three categories of libraries to validate certificates:
(1) the standard Android library,
(2) the Mbed TLS library~\cite{unity-mbedtls} provided by the Unity SDK, and
(3) the Unreal version of the OpenSSL library~\cite{unreal-openssl}.
\tool{} places Frida hooks into the certificate validation functions provided by these three libraries.
These hooks change the return value of the intercepted functions and set certain flags used to determine the validity of a certificate to ensure that AntMonitor's certificate is always trusted.
\minorrevision{
While bypassing certificate validation in the standard Android library is a widely known technique~\cite{frida-pinning-bypass}, bypassing validation in Unity and Unreal SDKs is not. Thus, we developed the following technique.
}


\myparagraph{Decrypting Unity and Unreal.}
Since most \device{} apps are developed using either the Unity or the Unreal game engines, they use the certificate validation functions provided by these engines instead of the ones in the standard Android library.
Below, we present our implementation of certificate validation bypassing for each engine.

For Unity, we discovered that the main function that is responsible for checking the validity of certificates is \code{mbedtls\_x509\_crt\_verify\_with\_profile()} in the Mbed TLS library, by inspecting its source code~\cite{mbedtls-cert-pinning}.
This library is used by the Unity framework as part of its SDK. Although Unity apps and its SDK are written in C\#, the final Unity library is a C++ binary.
When a Unity app is packaged for release, unused APIs and debugging symbols get removed from the Unity library's binary. 
This process makes it difficult to hook into Unity's functions since we cannot locate the address of a function of interest without having the symbol table to look up its address. Furthermore, since the binary also gets stripped of unused functions, we cannot rely on the debug versions of the binary to look up addresses because each app will have a different number of APIs included.
To address this challenge, \tool{} automatically analyzes the debug versions of the non-stripped Unity binaries (provided by the Unity engine), extracts the function signature (\ie a set of hexadecimal numbers) of \code{mbedtls\_x509\_crt\_verify\_with\_profile()}, and then looks for this signature in the stripped version of the binary to find its address. 
This address can then be used to create the necessary Frida hook for an app.
The details of this automated binary analysis can be found in \appdx{\appref{}~\ref{app:binary-analysis}}.

\mycomment{
For Unreal\footnote{Unreal apps oftentimes have OBB files (\ie graphics, media files, and large program assets) that need to be pushed back upon repackaging.},}
For Unreal, we discovered that the main function that is responsible for checking the validity of certificates is the function \code{x509\_verify\_cert()} in the OpenSSL library, which is integrated as part of the Unreal SDK.
Fortunately, the Unreal SDK binary file comes with a partial symbol table that contains the location of \code{x509\_verify\_cert()}, and thus, \tool{} can set a Frida hook for it.

\subsection{Network Traffic Dataset}
\label{sec:network-traffic-dataset}

\subsubsection{Raw Network Traffic Data}
\label{sec:network-traffic-dataset-raw}

\minorrevision{We used \tool{} to collect network traffic for 140\footnote{\minorrevision{The remaining 10 apps were excluded for the following reasons: (1) six apps could not be repackaged;
(2) two apps were browser apps, which would open up the web ecosystem, diverting our focus from VR; (3) one app was no longer found on the store---we created our lists of top apps 
one month ahead of our experiments;
and (4) one app could not start on the \device{} even without any of our modifications.}} apps in our corpus}
during the months of March and April 2021.
To exercise these 140 apps and collect their traffic, we manually interacted with each one for seven minutes.
Although there are existing tools that automate the exploration of regular (non-gaming) mobile apps (\eg \cite{li2017droidbot}), automatic interaction with a variety of games is an open research problem. 
Fortunately, manual testing allows us to customize app exploration 
and split our testing time between exploring menus within the app to cover more of the potential behavior, and actually playing the game, which better captures the typical usage by a human user. As shown by prior work, such testing criteria lead to more diverse network traffic and reveal more privacy-related data flows~\cite{jin2018mobipurpose,recon, varmarken2020tv}. 
Although our methodology might not be exhaustive, it is inline with prior work~\cite{mohajeri2019watching, varmarken2020tv}.

%
\tabref{}~\ref{tab:dataset-summary} presents the summary of our network traffic dataset.
We discovered 158 domains and 92 \esld{}s in \tcpflowtotal{} TCP flows that contain \packettotal{} packets.
Among the 140 apps, 96 were developed using the Unity framework, 31 were developed using the Unreal framework, and 13 were developed using other frameworks.
\vspace{-5pt}

\begin{table}[t!]
	\tabletextsize
	\begin{tabularx}{\columnwidth}{r | r | r r | r r }
	    \toprule
		\textbf{App Store} &  \textbf{ Apps} & 
		\textbf{Domains} & \textbf{\esld{s}} &
		\textbf{Packets} & \textbf{TCP Fl.}
		\\
		\midrule
	
		\topoculusfree & 43 & 85 & 48 & 2,818 & 2,126 \\
		
	    \topoculuspaid & 49 & 54 & 35 & 2,278 & 1,883 \\
	    
	    \topsidequest & 48 & 57 & 40 & 2,679 & 2,260 \\
	    \midrule
	    \rowcolor{brandeisblue!10}
	    \textbf{Total} & 140 & 158 & 92 & \packettotal{} & \tcpflowtotal{} \\
	    \bottomrule
	\end{tabularx}
	\vspace{-5pt}
	\caption{\textbf{Network traffic dataset summary.} Note that the same domains and eSLDs can appear across the three groups of ``App Store'', so their totals are based on unique counts.}
	\label{tab:dataset-summary}
	\vspace{-10pt}
\end{table}

\begin{figure*}[t!]
    \centering
    \begin{subfigure}{0.48\textwidth}
        \centering
        \includegraphics[width=\textwidth]{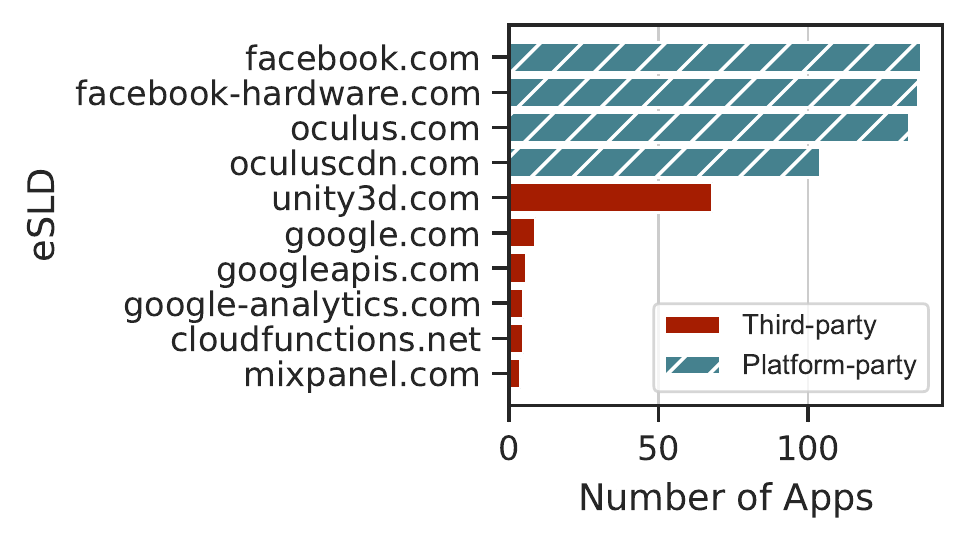}
        \vspace{-2em}
        \caption{}
        \label{fig:top-esld-third-and-platform-party}
    \end{subfigure}
    \hfill
    \begin{subfigure}{0.5\textwidth}
        \centering
        \includegraphics[width=\textwidth]{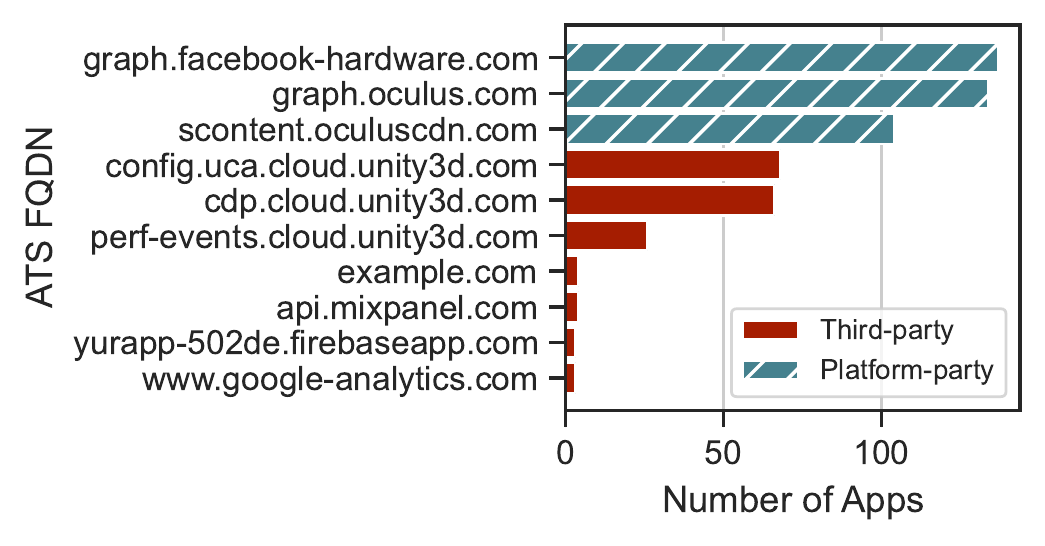}
        \vspace{-2em}
        \caption{}
        \label{fig:top-ats-third-and-platform-party}
    \end{subfigure}
    \vspace{-5pt}
    \caption{\textbf{Top-10 platform and third-party (a) eSLDs  and (b) ATS FQDNs.} They are ordered by the number of apps that contact them. Each app may have a few first-party domains: we found that 46 out of 140 (33\%) apps contact their own eSLDs.}
    \label{fig:top-esld-ats}
    \vspace{-10pt}
\end{figure*}

\subsubsection{Network Data Flows Extracted}
\label{sec:labeling-data-flow}
We processed the raw \netdataset{} and identified \dataflowtotal{} data flows: \dataflow{}. 
Next, we describe our methodology for extracting that information.

\myparagraph{App names.}
For each network packet, the app name is obtained by AntMonitor~\cite{shuba2016antmonitor}. This feature required a modification to work on Android 10, as described in
\appdx{\appref{}~\ref{app:antmonitor-improvement}}.



\myparagraph{Data types.}
The data types we extracted from our network traffic dataset are listed in \tabref{}~\ref{tab:pii-summary} and can be categorized into roughly three groups. 
First, we find personally identifiable information (\textit{PII}), including: user identifiers (\eg Name, Email, and User ID), device identifiers (Android ID, Device ID, and Serial Number), Geolocation, \etc 
Second, we found system parameters and settings, whose combinations are known to be used by trackers to create unique profiles of users~\cite{mohajeri2019watching, moz-fingerprinting}, \ie \textit{Fingerprints}. 
Examples include various version information (\eg{} Build and SDK Versions), Flags (\eg{} indicating whether the device is rooted or not), Hardware Info (\eg{} Device Model, CPU Vendor, \etc), Usage Time, \etc
Finally, we also find data types that are unique to VR devices (\eg{} VR Movement and VR Field of View) and group them as \textit{VR Sensory Data}. 
These can be used to uniquely identify a user or convey sensitive information---the VR Play Area, for instance, can represent the actual area of the user's household.

We use several approaches to find these data types in  HTTP headers and bodies, and also in any raw TCP segments that contain ASCII characters.
First, we use string matching to search for data that is static by nature. 
For example, we search for user profile data (\eg{} User Name, Email, \etc) using our test \oculusvr{} account and for any device identifiers (\eg{} Serial Number, Device ID, \etc) that can be retrieved by browsing the \device{} settings.
%
In addition, we search for their MD5 and SHA1 hashes.
Second, we utilize regular expressions to capture more dynamic data types. For example, we can capture different Unity SDK versions using \verb!UnityPlayer/[\d.]+\d!.
Finally, for cases where a packet contains structured data (\eg URL query parameters, HTTP Headers, JSON in HTTP body, \etc), we split the packet into key-value pairs and create a list of unique keys that appear in our entire network traffic dataset.
We then examine this list to discover keys that can be used to further enhance our search for data types.
For instance, we identified that the keys ``user\_id'' and ``x--playeruid'' can be used to find User IDs. 
\appdx{\appref{}~\ref{app:extracting-data-types-appendix}} provides more details on our data types. 
%


\myparagraph{Destinations.}
To extract the destination fully qualified domain name (FQDN), we use the HTTP Host field and the TLS SNI (for cases where we could not decrypt the traffic). 
Using tldextract, 
we also identify the effective second-level domain (eSLD) and use it to determine the high level organization that owns it via Crunchbase. 
We also adopt similar labeling methodologies from~\cite{varmarken2020tv} and~\cite{andow2020actions} to categorize each destination as either \emph{first-}, \emph{platform-}, or \emph{third-}party.
To perform the categorization, we also make use of collected privacy policies (see \figref{}~\ref{fig:our-work} and \secref{}~\ref{sec:privacy-policy}), as described next.
First, we tokenize the domain and the app's package name.
We label a domain as first-party if the domain's tokens either appear in the app's privacy policy URL or match the package name's tokens.
If the domain is part of cloud-based services (\eg \textit{vrapp.amazonaws.com}), we only consider the tokens in the subdomain (\textit{vrapp}). 
Second, we categorize the destination as platform-party if the domain contains the keywords ``oculus'' or ``facebook''.
Finally, we default to the third-party label.
This means that the data collection is performed by an entity that is not associated with app developers nor the platform, and the developer may not have control of the data being collected.
The next section presents further analysis of the destination domains.

\subsection{\oculusvr{} Advertising \& Tracking Ecosystem}
\label{sec:ats-ecosystem}
In this section, we explore the destination domains found in our network traffic dataset (see \secref{}~\ref{sec:labeling-data-flow}).
\figref{}~\ref{fig:top-esld-third-and-platform-party} presents the top-10 eSLDs for platform and third-party. 
We found that, unlike the mobile ecosystem, the presence of third-parties is minimal and platform traffic dominates in all apps (\eg{} $oculus.com$, $facebook.com$).
%
The most prominent third-party organization is Unity (\eg{} \textit{unity3d.com}), which appears in 68 out of 140 apps (49\%). 
This is expected since 96 apps in our dataset were developed using the Unity engine (see \secref{}~\ref{sec:network-traffic-dataset-raw}).
Conversely, although 31 apps in our dataset were developed using the Unreal engine, it does not appear as a major third-party data collector because Unreal does not provide its own analytics service. 
Beyond Unity, other small players include Alphabet (\eg{} \textit{google.com},  \textit{cloudfunctions.net}) and Amazon (\eg{} \textit{amazonaws.com}).
In addition, 87 out of 140 apps contact four or fewer third-party eSLDs (62\%). 
%

\myparagraph{Identifying ATS domains.} 
To identify ATS domains, we apply the following popular domain-based blocklists:
(1) \textit{PiHole's Default List}~\cite{pihole-homepage}, a list that blocks cross-platform ATS domains for IoT devices;
(2) \textit{Mother of All Adblocking}~\cite{moab}, a list that blocks both ads and tracking domains for mobile devices; and
(3) \textit{Disconnect Me}~\cite{disconnectme}, a list that blocks tracking domains.
For the rest of the paper, we will refer to the above lists simply as ``blocklists''.
We note that there are no blocklists that are curated for VR platforms. 
Thus, we choose blocklists that balance between IoT and mobile devices, and one that specializes in tracking.

\myparagraph{\oculusvr{} ATS ecosystem.} The majority of identified ATS domains relate to social and analytics-based purposes. \figref{}~\ref{fig:top-ats-third-and-platform-party} provides the top-10 ATS FQDNs that are labeled by our \blocklist{s}. 
We found that the prevalent platform-related FQDNs along with Unity, the prominent third party, are labeled as ATS. This is expected: domains such as \textit{graph.oculus.com} and \textit{perf-events.cloud.unity3d.com} are utilized for social features like managing leaderboards 
and app analytics, 
respectively.
We also consider the presence of organizations based on the number of unique domains contacted.
The most popular organization is Alphabet, which has 13 domains, such as \textit{google-analytics.com} and \textit{firebase-settings.crashlytics.com}. Four domains are associated with Facebook, such as \textit{graph.facebook.com}. Similarly, four are from Unity, such as \textit{userreporting.cloud.unity3d.com} and \textit{config.uca.cloud.unity3d.com}. 
Other domains are associated with analytics companies that focus on tracking how users interact with apps (\eg{} whether they sign up for an account) such as \textit{logs-01.loggly.com}, \textit{api.mixpanel.com}, and \textit{api2.amplitude.com}.
\mycomment{
The most popular organization is Alphabet, which has 13 domains, such as \textit{google-analytics}~\cite{google-analytics} and \textit{firebase-settings.crashlytics.com}~\cite{crashlytics}. Four domains are associated with Facebook, such as \textit{graph.facebook.com}~\cite{facebook-graph-api}. Similarly, four are from Unity, such as \textit{userreporting.cloud.unity3d.com} and \textit{config.uca.cloud.unity3d.com}\cite{unity-analytics}. 
Other domains are associated with analytics companies that focus on tracking how users interact with apps (\eg{} whether they sign up for an account) such as \textit{logs-01.loggly.com}~\cite{loggly}, \textit{api.mixpanel.com}~\cite{mixpanel}, and \textit{api2.amplitude.com}~\cite{amplitude-analytics}.}\minorrevision{Lastly, we provide an in-depth comparison to other ecosystems in~\secref{}~\ref{sec:major-findings}.}

\mycomment{
\myparagraph{\oculusvr{} vs. Android and Smart TV.} When comparing the \oculusvr{} ATS ecosystem with the more mature Android~\cite{shuba2020nomoats, razaghpanah2018apps} and Smart TVs (\eg{} Roku and Amazon's FireTV)~\cite{varmarken2020tv, mohajeri2019watching}, the biggest difference is that \oculusvr{} does not have discernible ad-related domains. For example, while Roku has \textit{ads.tremorhub.com}, FireTV has \textit{amazon-adsystem.com}, and Android has \textit{doubleclick.net}, \oculusvr{} only has their tracking counterparts. 
\rahmadi{We surmise that this is because Facebook wants all ads to be within its own ad ecosystem.}
Facebook's move to force new \oculusvr{} users to have a Facebook account in late 2020 points to this possibility~\cite{oculus-blog-force-login}. 
However, the \oculusvr{} ATS ecosystem does contain other major key-players, such as Unity and Alphabet, that are also present in the other two ecosystems. 
Still, Unity domains that are related to ads (\eg \textit{unityads.unity3d.com}~\cite{razaghpanah2018apps}) do not appear in the \oculusvr{} ATS ecosystem. 
Furthermore, the \oculusvr{} ATS ecosystem is not as diverse---it misses big companies like Amazon, Comscore, Inc (\eg{} \textit{scorecardresearch.com}) and Adobe (\eg{} \textit{dpm.demdex.net})~\cite{varmarken2020tv}. 
}

\begin{table}[t!]
	\tabletextsize
	\centering
	\begin{tabularx}{\linewidth}{p{3.9cm} p{2cm} p{1.4cm} }
	    \toprule
		\textbf{FQDN} & \textbf{Organization} & \textbf{Data Types} 
		\\
		\midrule
		bdb51.playfabapi.com & Microsoft & 11 \\
		\midrule
		sharedprod.braincloudservers.com & bitHeads Inc. & 8   \\
		\midrule
		cloud.liveswitch.io & Frozen Mountain Software & 7 \\
		\midrule
		datarouter.ol.epicgames.com & Epic Games & 6 \\
		\midrule
		9e0j15elj5.execute-api.us-west-1.amazonaws.com & Amazon & 5  \\ 
	    \bottomrule
	\end{tabularx}
	\caption{Top-5 third-party FQDNs that are missed by blocklists based on the number of data types exposed.}
	\vspace{-1em}
	\label{tab:missed-fqdns}
\end{table}

\myparagraph{Missed by blocklists.} 
The three \blocklist{s} that we use in \tool{} are not tailored for the Oculus platform. As a result, there could be domains that are ATS related but not labeled as such. To that end, we explored and leveraged data flows to find potential domains that are missed by \blocklist{s}. In particular, we start from data types exposed in our network traffic, and identify the destinations where these data types are sent to.  \tabref{}~\ref{tab:missed-fqdns} summarizes third-party destinations that collect the most data types and are {\em not} already captured by any of the \blocklist{s}.
We found the presence of 11 different organizations, not caught by blocklists, including: Microsoft, bitHeads Inc., and Epic Games---the company that created the Unreal engine. 
The majority are cloud-based services that provide social features, such as messaging, and the ability to track users for engagement and monetization (\eg{} promotions to different segments of users). 
We provide additional FQDNs missed by blocklists in \appdx{\appref{}~\ref{app:missed-blocklists}}.

\mycomment{
We found the presence of 11 different organizations, not caught by blocklists, including: Microsoft~\cite{microsoft-playfab}, bitHeads Inc.~\cite{bitheads}, and Epic Games (the company that created the Unreal engine)~\cite{epicgames}. The majority are cloud-based services that provide social features, such as messaging, and the ability to track users for engagement and monetization (\eg{} promotions to different segments of users)~\cite{azure-business-intel, braincloud-features, gamesparks}. We provide additional FQDNs missed by blocklists in \appref{}~\ref{app:missed-blocklists}.}

\begin{table}[t!]
	\tabletextsize
	\centering
	\minorrevision{
	\begin{tabularx}{\columnwidth}{ p{18mm} | r r r | r r r | r r r}
	    \toprule
		\multicolumn{1}{l|}{Data Types (21)}  & \multicolumn{3}{c|}{\textbf{Apps}} &  \multicolumn{3}{c}{\textbf{FQDNs}} &  \multicolumn{3}{c}{\textbf{\% Blocked}}  \\
		\multicolumn{1}{l|}{\textbf{PII}} & 
		\multicolumn{1}{X}{1\textsuperscript{st}} & 
		\multicolumn{1}{X}{3\textsuperscript{rd}} & 
		\multicolumn{1}{X|}{Pl.} & 
		\multicolumn{1}{X}{1\textsuperscript{st}} & 
		\multicolumn{1}{X}{3\textsuperscript{rd}} & 
		\multicolumn{1}{X|}{Pl.} & 
		\multicolumn{1}{X}{1\textsuperscript{st}} & 
		\multicolumn{1}{X}{3\textsuperscript{rd}} & 
		\multicolumn{1}{X}{Pl.} 
		\\
        
		\midrule
        Device ID  & 6 & 64 & 2  & 6 & 13 & 1 & 0 & 38 & 100 \\
		User ID  & 5 & 65 & 0 & 5 & 13 & 0 & 20 & 38 & - \\
        Android ID & 6 & 31 & 18 & 6 & 7  & 2 & 17 & 43 & 50 \\
		Serial Number & 0 & 0 & 18 & 0 & 0 & 2 & - & - & 50  \\
		Person Name & 1 & 7 & 0 & 1 & 4 & 0 & 0 & 50 & - \\
        Email  & 2 & 5 & 0 & 2 & 5 & 0 & 0 & 20 & - \\
	    Geolocation & 0 & 5 & 0 & 0 & 4 & 0 & - & 50 & - \\
		\midrule	
		\multicolumn{10}{l}{\textbf{Fingerprint}} \\
		\midrule
        SDK Version & 23 & 69 & 20 & 34 & 28 & 4 & 6 & 46 & 0 \\
        Hardware Info & 21 & 65 & 19 & 25 & 23 & 3 & 4 & 39 & 33 \\
        System Version & 16 & 62 & 19 & 20 & 21 & 3 & 5 & 43 & 33 \\
        Session Info  & 7 & 66 & 2 & 7 & 13 & 1 & 14 & 46 & 100 \\
        App Name  & 4 & 65 & 2 & 4 & 10 &1 & 25 & 40 & 100 \\
        Build Version  & 0 & 61 & 0 & 0 & 3 & 0 & - & 100 & -  \\
        Flags  & 6 & 53 & 2 & 6 & 8 & 1 & 0 & 50 &100 \\
        Usage Time & 2 & 59 & 0 & 2 & 4 & 0 & 0 & 50 & - \\        
        Language  & 5 & 28 & 16 & 5 & 9 & 1 & 0 & 56 & 0 \\
        Cookies  & 5 & 4 & 2 & 5 & 3 & 1 & 0 & 33 & 100 \\
        \midrule
        \multicolumn{10}{l}{\textbf{VR Sensory Data}} \\ 
        \midrule
        VR Play Area  & 0 & 40 & 0 & 0 & 1 & 0 & - & 100 & - \\
        VR Movement  & 1 & 24 & 2 & 1 & 6 & 1 & 0 & 67 & 100 \\
        VR~Field~of~View & 0 & 16 & 0 & 0 & 1 & 0 & - & 100 & - \\
        VR Pupillary & 0 & 16 & 0 & 0 & 1 & 0 & - & 100 & - \\
        Distance & & & & & & & & & \\

       \midrule
       \rowcolor{brandeisblue!10}
        \textbf{Total} & 33 & 70 & 22 & 44 & 39 & 5 & 5 & 36 & 20 \\
	    \bottomrule
	\end{tabularx}
		\vspace{-5pt}
		}
	\caption{\textbf{Data types exposed in the network traffic dataset.}
	\minorrevision{Column ``Apps'' reports the number of apps that send the data type to a destination; column ``FQDNs'' reports the number of FQDNs that receive that data type; and column ``\% Blocked'' reports the percentage of FQDNs blocked by \blocklist{s}. Using sub-columns, we denote party categories: first (1\textsuperscript{st}), third (3\textsuperscript{rd}), and platform (Pl.) parties. 
	}
	}
	\label{tab:pii-summary}
\end{table}

\subsection{Data Flows in Context}
\label{sec:pii-exposures}

The exposure of a particular data type, on its own, does not convey much information: it may be appropriate or inappropriate depending on the context~\cite{privacy-in-context}. For example, geolocation sent to the GoogleEarth VR or Wander VR app is necessary for the  functionality, while geolocation used for ATS purposes is less appropriate. The network traffic can be used to partly infer the purpose  of data flows, \eg depending on whether the destination being first-, third-, or platform-party; or an ATS.
\tabref{}~\ref{tab:pii-summary} lists all data types found in our network traffic, extracted using the methods explained in \secref{}~\ref{sec:labeling-data-flow}. 


\myparagraph{Third party.} 
Half of the apps (70 out of \appstotal) expose data flows to third-party FQDNs, 36\% of which are labeled as ATS by blocklists.
Third parties collect a number of PII data types, including Device ID (64 apps), User ID (65 apps), and Android ID (31 apps), indicating cross-app tracking.
In addition, third parties collect system, hardware, and version info from over 60 apps---denoting the possibility that the data types are utilized to fingerprint users.
Further, all VR specific data types, with the exception of VR Movement, are collected by a single third-party ATS domain belonging to Unity.
VR Movement is collected by a diverse set of third-party destinations, such as \textit{google-analytics.com}, \textit{playfabapi.com} and \textit{logs-01.loggly.com}, implying that trackers are becoming interested in collecting VR analytics. 

\myparagraph{Platform party.} 
Our findings on exposures to platform-party domains are a lower bound since not all platform traffic could be decrypted (see \secref{}~\ref{sec:limitation-conclusion}). 
However, even with limited decryption, we see a number of exposures whose destinations are five third-party FQDNs. 
Although only one of these FQDNs is labeled as ATS by the blocklists, other platform-party FQDNs could be ATS domains that are missed by blocklists (see \secref{}~\ref{sec:ats-ecosystem}). 
For example, \textit{graph.facebook.com} is an ATS FQDN, and \textit{graph.oculus.com} appears to be its counterpart for \oculusvr{}; it collects six different data types in our dataset. 
Notably, the platform party is the sole party responsible for collecting a sensitive hardware ID that cannot be reset by the user---the Serial Number.
In contrast to OVR, the Android developer guide strongly discourages its use~\cite{android-dev-practice}.

\myparagraph{First party.} 
Only 33 apps expose data flows to first-party FQDNs, and only 5\% of them are labeled as ATS.
Interestingly, the blocklists tend to have higher block rates for first-party FQDNs if they collect certain data types, \eg Android ID (17\%), User ID (20\%), and App Name (25\%).
Popular data types collected by first-party destinations are Hardware Info (21 apps), SDK Version (23 apps), and System Version (16 apps). For developers, this information can be used to prioritize bug fixes or improvements that would impact the most users. Thus, it makes sense that only \textasciitilde5\% of first-party FQDNs that collect this information are labeled as ATS.
\myparagraph{Summary.}
The \oculusvr{} ATS ecosystem is young when compared to Android and Smart TVs. It is dominated by tracking domains for social features and analytics, but not by ads. We have detailed 21 different data types that \oculusvr{} sends to first-, third-, and platform-parties. State-of-the-art blocklists only captured 36\% of exposures to third parties, missing some sensitive exposures such as Email, User ID, and Device ID.

\section{\tool{}: Privacy Policy Analysis}
\label{sec:privacy-policy}

In this section, we turn our attention to the intended data collection and sharing practices, as stated in the text privacy policy. 
For example, from the text {\em ''We may collect your email address and share it for advertising purposes''}, we want to extract the collection statement (``we'', which implies the app's first-party entity; ``collect'' as action; and ``email address'' as data type) and the purpose (``advertising''). In \secref{}~\ref{sec:policheck-polisys}, we present our methodology for extracting data collection statements, and comparing them against data flows found in network traffic for consistency. \tool{} builds and improves on state-of-the-art NLP-based tools: \policheck{}~\cite{andow2020actions} and \policylint{}~\cite{andow2019policylint}, previously developed for mobile apps. In \secref{}~\ref{sec:policy-ontology}, we present our VR-specific ontologies for data types and entities.  
In \secref{}~\ref{sec:policheck-results}, we report network-to-policy consistency results.
\secref{}~\ref{sec:polisis-context} describes how we interface between the different NLP models of \policheck{} and  \polisys{} to extract the data collection purpose and other context for each data flow. 

\myparagraph{Collecting privacy policies.}
For each app in \secref{}~\ref{sec:network-traffic-analysis}, we also collected its privacy policy on the same day that we collected its network traffic. 
Specifically, we used an automated Selenium~\cite{selenium} script to crawl the webstore and extracted URLs of privacy policies. For apps without a policy listed, we followed the link to the developer's website to find a privacy policy.  We also included eight third-party policies (\eg from Unity, Google), referred to by the apps' policies.

For the top-50 free apps on the Oculus store, we found that only 34 out of the 43 apps have privacy policies.  Surprisingly, for the top-50 paid apps, we found that only 39 out of 49 apps have privacy policies.
For the top-50 apps on SideQuest, we found that only 29 out of 48 apps have privacy policies.
Overall, among apps in our corpus, we found that only 102 (out of 140) apps provide valid English privacy policies.   We treated the remaining apps as having empty privacy policies, ultimately leading \tool{} to classify their data flows as omitted disclosures.
\vspace{-5pt}


\subsection{Network-to-Policy Consistency}
\label{sec:policheck-improvement}

Our goal is to analyze text in the app's privacy policy, extract statements about data collection (and sharing), and compare them against the actual data flows found in network traffic. 

\subsubsection{Consistency Analysis System}
\label{sec:policheck-polisys}

\tool{} builds on state-of-the-art tools: \policylint{}~\cite{andow2019policylint} and \policheck{}~\cite{andow2020actions}.  \policylint{}~\cite{andow2019policylint} provides an NLP pipeline that takes a sentence as input. For example, it takes the sentence {\em ``We may collect your email address and share it for advertising purposes''}, and extracts the collection statement ``(entity: \emph{we}, action: \emph{collect}, data type: \emph{email address})''.
More generally, \policylint{} takes the app's privacy policy text, parses sentences and performs standard NLP processing, and eventually extracts data collection statements defined as the tuple $P=$\polichecktuple{}, where \textit{app} is the sender and \textit{entity} is the recipient performing an \textit{action} (collect or not collect) on the \textit{data type}.
\policheck{}~\cite{andow2020actions}  takes the app's data flows (extracted from the network traffic and defined as $F=$\policheckdataflow{}) and compares it against the stated $P$ for consistency. 

\policheck{} classifies the disclosure of $F$ as {\em clear} (if the data flow exactly matches a collection statement), {\em vague} (if the data flow  matches a collection statement in broader terms), {\em omitted} (if there is no collection statement corresponding to the data flow), {\em ambiguous} (if there are contradicting collection statements about a data flow), or {\em incorrect} (if there is a data flow for which the collection statement states otherwise). 
Following \policheck{}'s terminology~\cite{andow2020actions}, we further group these five types of disclosures into  two groups: \emph{consistent} (clear and vague disclosures) and \emph{inconsistent} (omitted, ambiguous, and incorrect) disclosures. 
The idea is that for consistent disclosures,  there is a statement in the policy that matches the data type and entity, either clearly or vaguely.
\minorrevision{
 \tabref{}~\ref{tab:policy-sentence-vs-data-flow} provides real examples of data collection disclosures extracted from VR apps that we analyzed.
}

\begin{table*}[t!]
	\tabletextsize
	\centering
	\minorrevision{
	\begin{tabularx}{\linewidth}{c | p{15mm} p{55mm} p{50mm} p{35mm}}
	    \toprule
		\multicolumn{2}{c}{\textbf{Disclosure Type}} &  \textbf{Privacy Policy Text} & 
		\textbf{Action : Data Collection Statement (P)} & \textbf{Data Flow (F)} 
		\\
		\midrule
		\multirow{2}{*}{\rotatebox[origin=c]{90}{\textbf{Consistent}}}
		& Clear & ``For example, we collect information ..., and \textit{a timestamp for the request}.'' & \emph{collect : $\langle$com.cvr.terminus, usage time, we$\rangle$} & \emph{$\langle$usage time, we$\rangle$} \\
		\cline{2-5}
	    & Vague & ``We will share \emph{your information (in some cases} & \emph{collect : $\langle$com.HomeNetGames.WW1oculus,} & \emph{$\langle$serial number, oculus$\rangle$}\\
	    & & \emph{personal information)} with third-parties, ...'' & \emph{pii, third party$\rangle$} & \emph{$\langle$android id, oculus$\rangle$}\\
	    \midrule
	    & Omitted & - & \emph{collect : $\langle$com.kluge.SynthRiders, -, -$\rangle$} & \emph{$\langle$system version, oculus$\rangle$}\\
	    & & & & \emph{$\langle$sdk version, oculus$\rangle$}\\
	    & & & & \emph{$\langle$hardware information, oculus$\rangle$}\\
	    \cline{2-5}
	    \multirow{3}{*}{\rotatebox[origin=c]{90}{\textbf{Inconsistent}}}
	    & Ambiguous & ``..., Skydance \emph{will not disclose any Personally} & \emph{collect : $\langle$com.SDI.TWD, pii, third party$\rangle$} & \emph{$\langle$serial number, oculus$\rangle$}\\
	    & & \emph{Identifiable Information to third parties} ... & & \emph{$\langle$android id, oculus$\rangle$}\\
	    & & your \emph{Personally Identifiable Information will be disclosed to such third parties} and ...'' & & \\
	    \cline{2-5}
	    & Incorrect & ``We \emph{do not share our customer's personal in-} & \emph{not\_collect : $\langle$com.downpourinteractive.} & \emph{$\langle$device id, unity$\rangle$}\\
	    & & \emph{formation with unaffiliated third parties} ...'' & \emph{onward, pii, third party$\rangle$} & \emph{$\langle$user id, oculus$\rangle$}\\
	    \bottomrule
	\end{tabularx}
    }
	\vspace{-5pt}
	\caption{\minorrevision{\textbf{Examples to illustrate the types of disclosures identified by \policheck{}.} 
A {\em data collection statement (P)} is extracted from the privacy policy text and is defined as the tuple $P=$\polichecktuple{}. {\em A data flow (F)} is extracted from the network traffic and is defined as $F=$\policheckdataflow{}. During the consistency analysis, each $P$ can be mapped to zero, one, or more $F$.}}
	\label{tab:policy-sentence-vs-data-flow}
	\vspace{-10pt}
\end{table*}

\begin{figure*}
    \centering
    \vspace{-5pt}
    \begin{subfigure}[b]{0.48\textwidth}
        \centering
        \includegraphics[width=\textwidth]{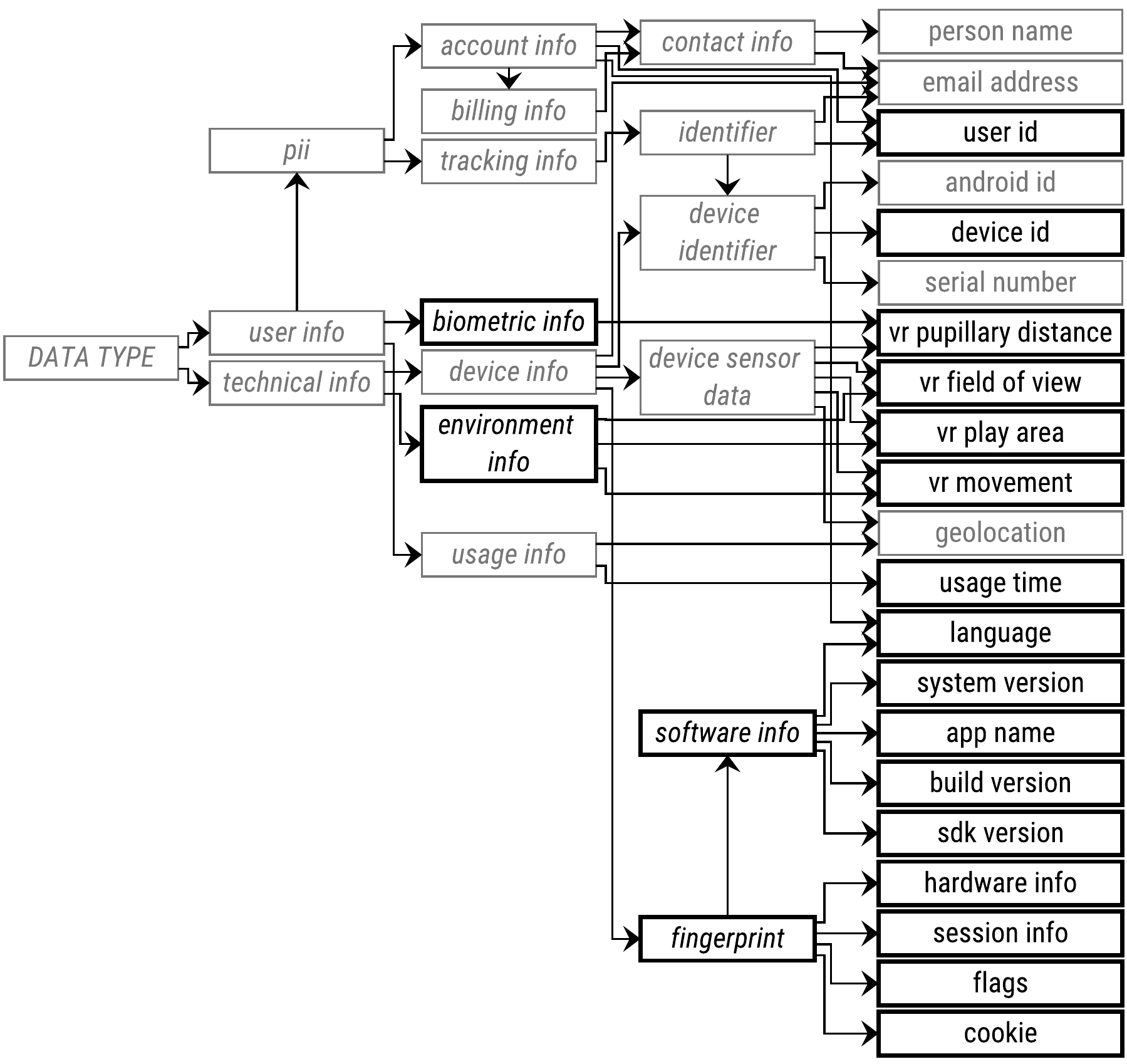}
        \caption{Data Ontology}
        \label{fig:data-ontology}
    \end{subfigure}
    \hfill
    \begin{subfigure}[b]{0.39\textwidth}
        \centering
        \includegraphics[width=\textwidth]{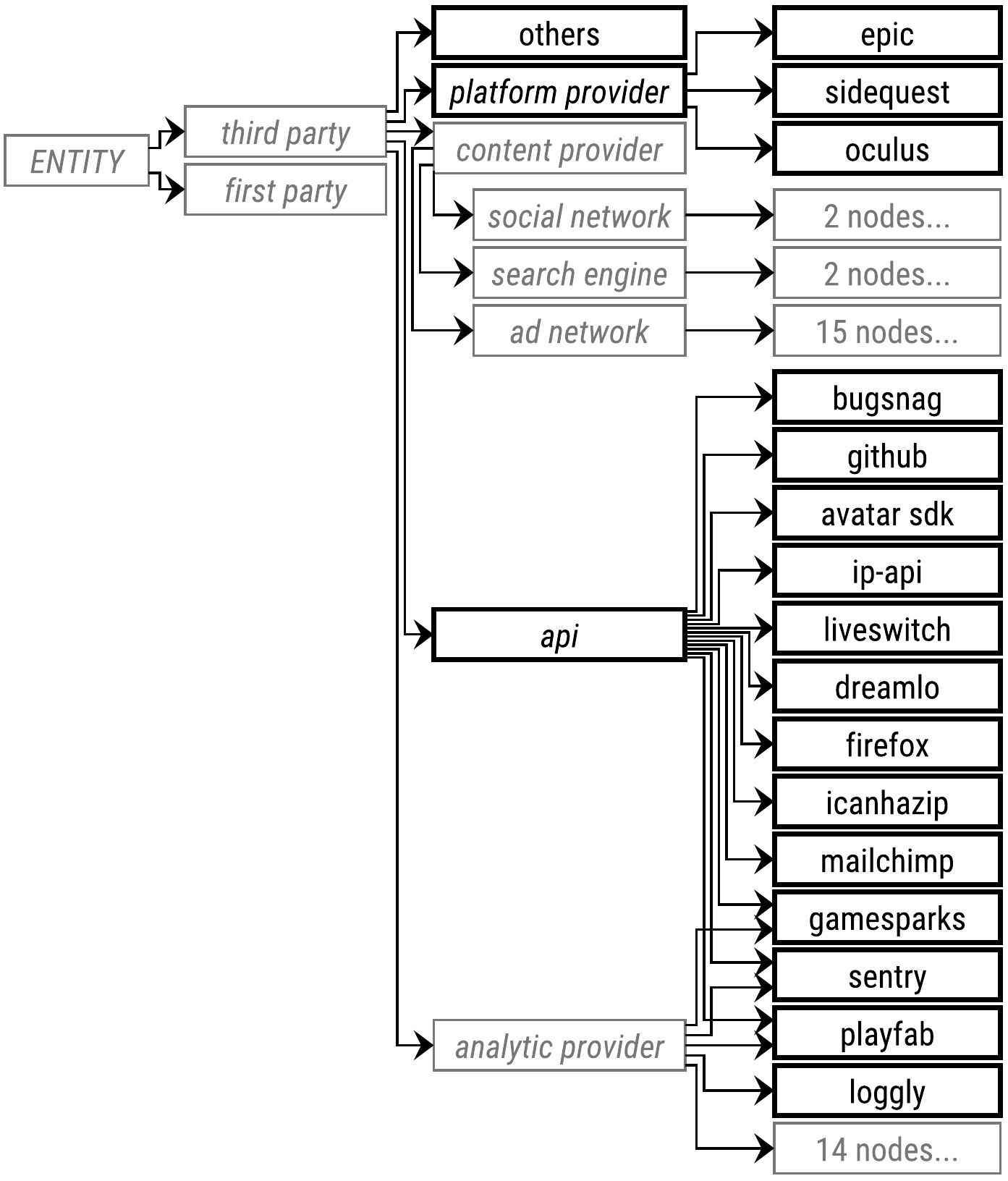}
        \caption{\minorrevision{Entity Ontology}}
        \label{fig:entity-ontology}
    \end{subfigure}
    \vspace{-8pt}
    \caption{\minorrevision{\textbf{Ontologies for VR data flows.} Please recall that each data flow, {\em F}, is defined as $F=$\policheckdataflow{}. We started from the \policheck{} ontologies, originally developed for Android (printed in gray). First, we eliminated nodes that did not appear in our VR network traffic and privacy policies. Then, we added new leaf nodes (printed in black) based on new data types found in the VR network traffic and/or privacy policies text. Finally, we defined additional non-leaf nodes, such as  ``biometric info'' and ''api'', in the resulting VR data and entity ontologies.}}
    \vspace{-1em}
    \label{fig:ontologies}
\end{figure*}

Consistency analysis relies on pre-built ontologies and synonym lists used to match (i) the data type and destination that appear in each $F$ with (ii) any instance of $P$ that discloses the same (or a broader) data type and destination\footnote{For example (see \figref{}~\ref{fig:data-ontology}), ``email address'' is a special case of ``contact info'' and, eventually, of ``pii''. There is a clear disclosure \wrt data type if the ``email address'' is found in a data flow and a collection statement. A vague disclosure is declared if the ``email address'' is found in a data flow and a collection statement that uses the term ``pii'' in the privacy policy. An omitted disclosure means that ``email address'' is found in a data flow, but there is no mention of it (or any of its broader terms) in the privacy policy.}.
 \tool{}'s adaptation of ontologies specifically for VR is described in  \secref{}~\ref{sec:policy-ontology}. 
We also improved several aspects of \policheck{}, as described in detail in \appdx{\appref{}~\ref{app:other-policheck-improvements}}.
First, we added a feature  to include a third-party privacy policy for analysis if it is mentioned in the app's policy. We found that 30\% (31/102) of our apps' privacy policies reference third-party privacy policies, and the original \policheck{} would mislabel third-party data flows from these apps as omitted. Second, we added a feature to more accurately resolve first-party entity names. Previously, only first-person pronouns (\eg ``we'') were used to indicate a first-party reference, while some privacy policies use company and app names in first-party references. The original \policheck{} would incorrectly recognize these first-party references as third-party entities for 16\% (16/102) of our apps' privacy policies.

\subsubsection{Building Ontologies for VR}
\label{sec:policy-ontology}

Ontologies are used to represent subsumptive relationships between terms: a link from term \textit{A} to term \textit{B} indicates that \textit{A} is a broader term (\textit{hypernym}) that subsumes \textit{B}.  
There are two ontologies, namely data and entity ontologies: the data ontology maps data types and entity ontology maps destination entities. 
 Since \policheck{} was originally designed for Android mobile app's privacy policies, it is important to adapt the ontologies to include data types and destinations specific to VR's privacy policies and  actual data flows.

\myparagraph{VR data ontology.} 
\figref{}~\ref{fig:data-ontology} shows the data ontology we developed for VR apps.
Leaf nodes correspond to all 21 data types found in the network traffic and listed in \tabref{}~\ref{tab:pii-summary}. Non-leaf nodes are broader terms extracted from privacy policies and may subsume more specific data types, \eg ``device identifier'' is a non-leaf node that subsumes ``android id''. We built a VR data ontology, starting from  the original Android data ontology, in a few steps as follows.
First, we cleaned up the original data ontology by removing data types that do not exist on \oculusvr{} (\eg ``IMEI'', ``SIM serial number'', \etc). We also merged similar terms (\eg ``account information'' and ``registration information'') to make the structure clearer. 
\minorrevision{
Next, we used PoliCheck to parse privacy policies from VR apps.
When PoliCheck parses the sentences in a privacy policy, it extracts terms and tries to match them with the nodes in the data ontology and the synonym list. If PoliCheck does not find a match for the term, it will save it in a log file.
We inspected each term from this log file, and added it either as a new node in the data ontology or as a synonym to an existing term in the synonym list.
}
Finally, we added new terms for data types identified in network traffic (see \secref{}~\ref{sec:pii-exposures}) as leaf nodes in the ontology. 
Most notably, we added VR-specific data types (see VR Sensory Data category shown in \tabref{}~\ref{tab:pii-summary}): ``biometric info'' and ``environment info''. The term ``biometric info'' includes physical characteristics of human body (\eg height, weight, voice, \etc); we found some VR apps that collect user's ``pupillary distance'' information. The term ``environment information'' includes VR-specific sensory information that describes the physical environment; we found some VR apps that collect user's ``play area'' and ``movement''.
\tabref{}~\ref{tab:new-ontology-summary} shows the summary of the new VR data ontology. It consists of 63 nodes: 39 nodes are new in \tool{}'s data ontology.
Overall, the original Android data ontology was used to track 12 data types (\ie 12 leaf nodes)~\cite{andow2020actions}, whereas our VR data ontology is used to track 21 data types (\ie 21 leaf nodes) appearing in the network traffic (see~\tabref{}~\ref{tab:pii-summary} and \figref{}~\ref{fig:data-ontology}).

\myparagraph{VR entity ontology.} 
Entities are names of companies and other organizations which refer to destinations. 
\minorrevision{
We use a list of domain-to-entity mappings to determine which entity  each domain belongs to (see \appdx{\appref{}~\ref{app:other-policheck-improvements}})---domain extraction and categorization as either first-, third-, or platform-party are described in detail in \secref{}~\ref{sec:labeling-data-flow}.} 
We modified the Android entity ontology to adapt it to VR as follows: (1) 
we pruned entities that were not found in privacy policies of VR apps or in our network traffic dataset, and (2) we added new entities found in both sources.
\tabref{}~\ref{tab:new-ontology-summary} summarizes the new entity ontology. It consists of 64 nodes: 21 nodes are new in \tool{}'s entity ontology.
\figref{}~\ref{fig:entity-ontology} shows our VR entity ontology, in which we added two new non-leaf nodes: ``platform provider'' (which includes online distribution platforms or app stores that support the distribution of VR apps)  and ``api'' (which refers to various third-party APIs and services that do not belong to existing entities).
We identified 16 new entities that were not included in the original entity ontology. We visited the websites of those new entities and found that: three are platform providers, four are analytic providers, and 12 are service providers; these become the leaf nodes of ``api''.
We also added a new leaf node called ``others'' to cover a few data flows, whose destinations cannot be determined from the IP address or domain name.

\begin{table}[t!]
	\centering
	\begin{tabularx}{\linewidth}{X  r  r}
	    \toprule
		\textbf{Platform} & \textbf{Data Ontology} & \textbf{Entity Ontology}
		\\
		\midrule
		Android ~\cite{andow2020actions} & 38 nodes & 209 nodes \\
		\oculusvr{} (\tool) & 63 nodes & 64 nodes \\
		\midrule
		\rowcolor{brandeisblue!10}
		\textit{New nodes in \oculusvr{}} & 39 nodes            &  21 nodes  \\
	    \bottomrule
	\end{tabularx}
	\caption{Comparison of \policheck{} and \tool{} Ontologies. Nodes include leaf nodes (21 data types 
	 and 16 entities) and non-leaf nodes (see \figref{}~\ref{fig:ontologies}). }
	\label{tab:new-ontology-summary}
	\vspace{-12pt}
\end{table}

\myparagraph{Summary.}
Building VR ontologies has been non-trivial.
We had to examine a list of more than 500 new terms and phrases that were not part of the original ontologies.
Next, we had to decide whether to add a term into the ontology as a new node, or as a synonym to an existing node.
In the meantime, we had to remove certain nodes irrelevant to VR and merge others because the original Android ontologies were partially machine-generated and not carefully curated.

\begin{figure*}[!t]
	\centering
	\vspace{-10pt}
	\includegraphics[width=0.9\textwidth]{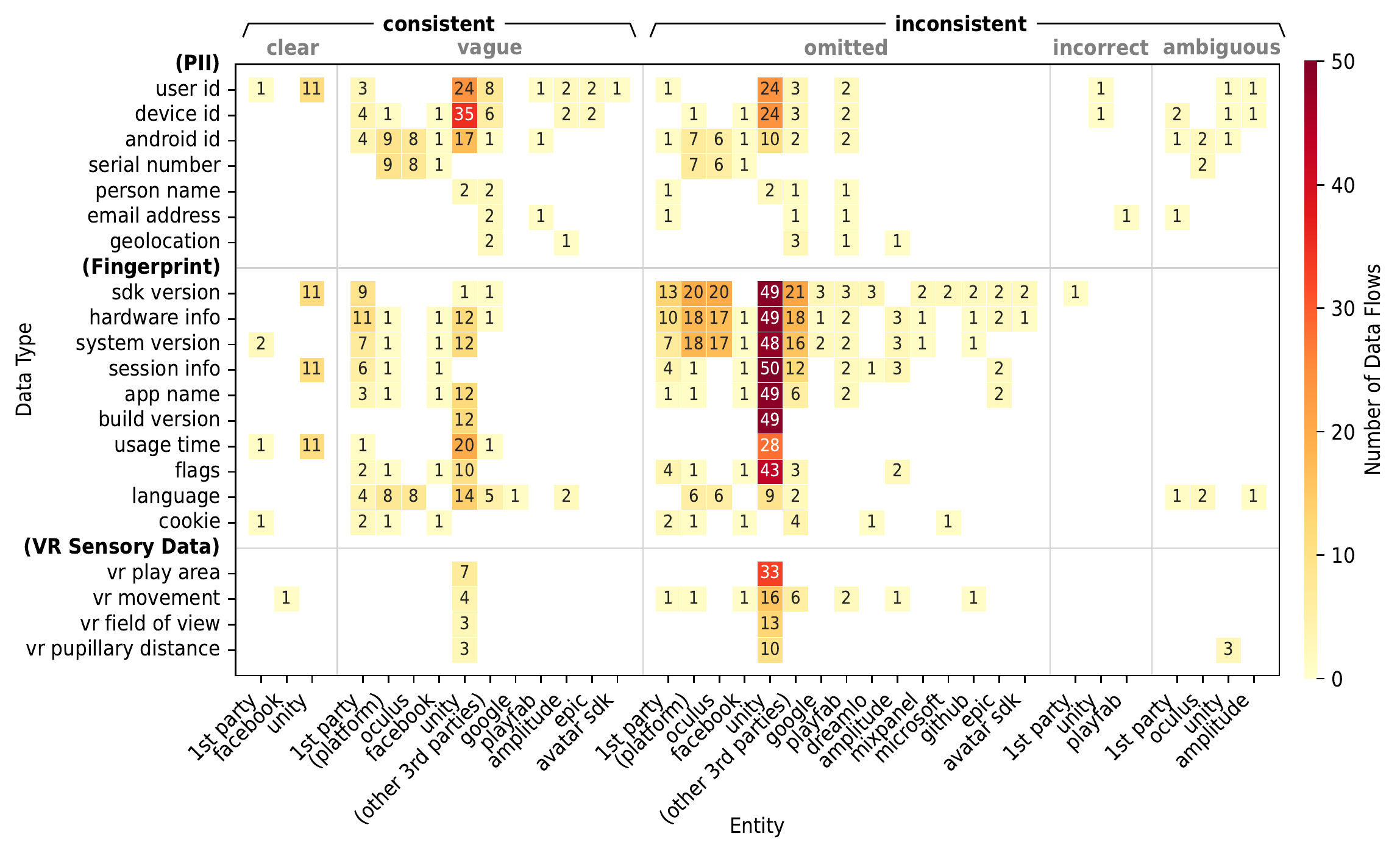}
	\vspace{-1em}
	\caption{\textbf{Summary of network-to-policy consistency analysis results.} Columns whose labels are in parentheses provide aggregate values: \eg column ``(platform)'' aggregates the columns ``oculus'' and ``facebook''; column ``(other 3rd parties)'' aggregates the subsequent columns. The numbers count data flows; each data flow is defined as \dataflowtuple{}).}
	\label{fig:heatmap-all}
	\vspace{-10pt}
\end{figure*}

\subsubsection{Network-to-Policy Consistency Results}

\label{sec:policheck-results}
We ran \tool{}'s privacy policy analyzer to perform network-to-policy consistency analysis. Please recall that we extracted \dataflowtotal{} data flows from 140 apps (see \secref{}~\ref{sec:labeling-data-flow}).  


\myparagraph{\oculusvr{} data flow  consistency.}
In total, 68\% (776/1,135) data flows are classified as inconsistent disclosures.  The large majority of them 97\% (752/776) are omitted disclosures, which are not declared at all in the apps' respective privacy policies.
\figref{}~\ref{fig:heatmap-all} presents the data-flow-to-policy consistency analysis results. Out of 93 apps which expose data types, 82 apps have at least one inconsistent data flows. 
Among the remaining 32\% (359/1,135) consistent data flows, 86\% (309/359) are classified as vague disclosures. They are declared in vague terms in the privacy policies (\eg the app's data flows contain the data type ``email address'', whereas its privacy policy only declares that the app collects ``personal information''). Clear disclosures are found in only 16 apps.



\begin{figure}[t!]
    \centering
    \begin{subfigure}[t!]{0.4\textwidth}
        \centering
        \includegraphics[width=\textwidth]{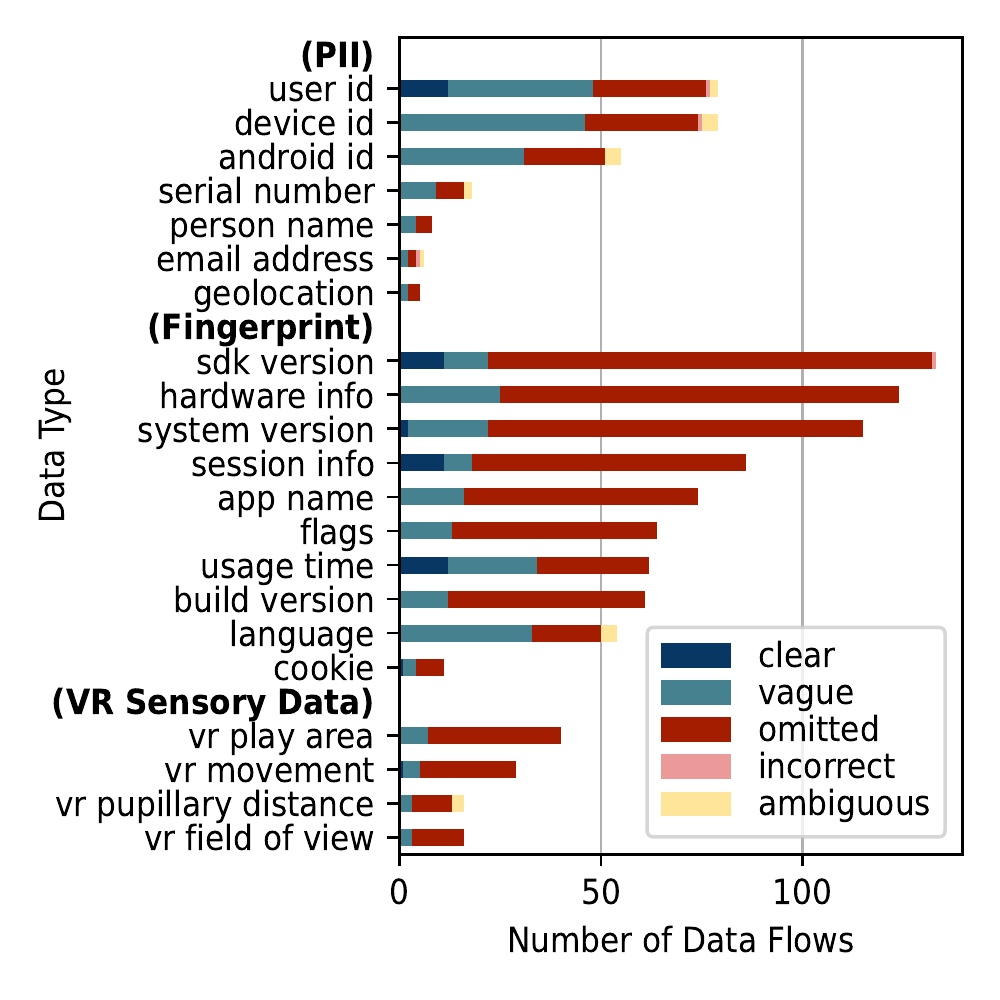}
        \vspace{-2.5em}
        \caption{}
        \label{fig:policheck-bar-dtype-5types}
    \end{subfigure}
    \hfill\\
    \begin{subfigure}[t!]{0.4\textwidth}
        \centering
        \includegraphics[width=\textwidth]{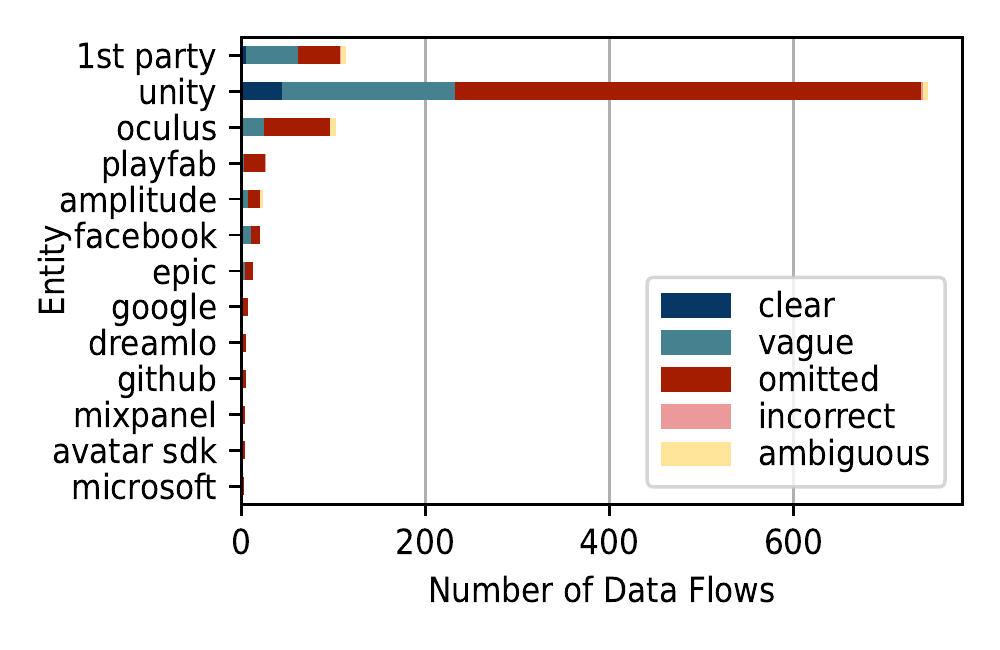}
        \vspace{-2.5em}
        \caption{}
        \label{fig:policheck-bar-entity-5types}
    \end{subfigure}
    \vspace{-5pt}
    \caption{Network-to-policy consistency analysis results aggregated by (a) data types, and (b) destination entities.}
    \label{fig:flow-to-policy-consistency-analysis}
    \vspace{-10pt}
\end{figure}

\mycomment{
\begin{figure}[ht!]
	\centering
	\includegraphics[width=1\columnwidth]{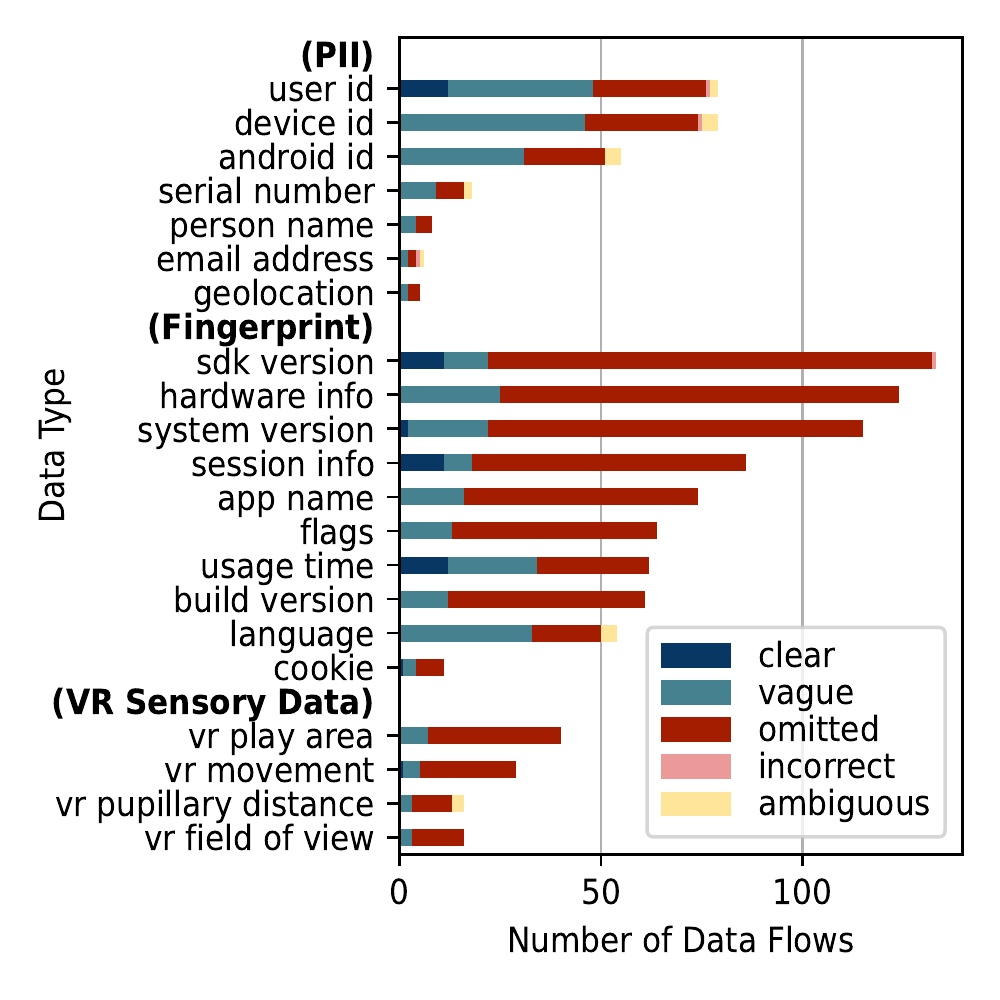}
	\caption{\term{Network}-to-policy consistency analysis results aggregated by data types.}
	\label{fig:policheck-bar-dtype-5types}
\end{figure}

\begin{figure}[ht!]
	\centering
	\includegraphics[width=1\columnwidth]{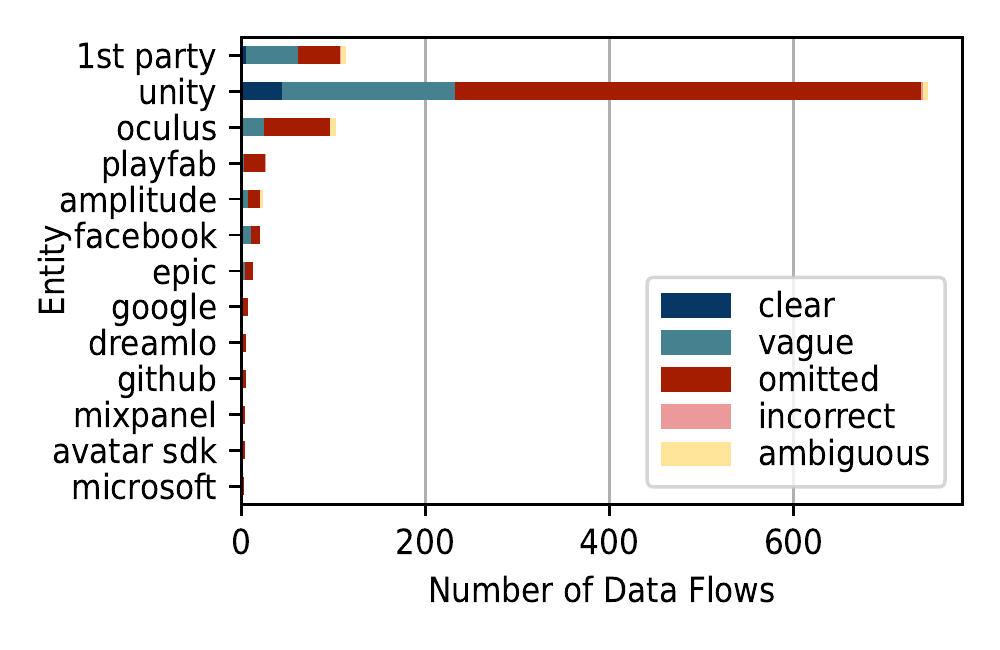}
	\caption{\term{Network}-to-policy consistency analysis results aggregated by destination entities.}
	\label{fig:policheck-bar-entity-5types}
\end{figure}

\begin{figure}[ht!]
	\centering
	\includegraphics[width=1\columnwidth]{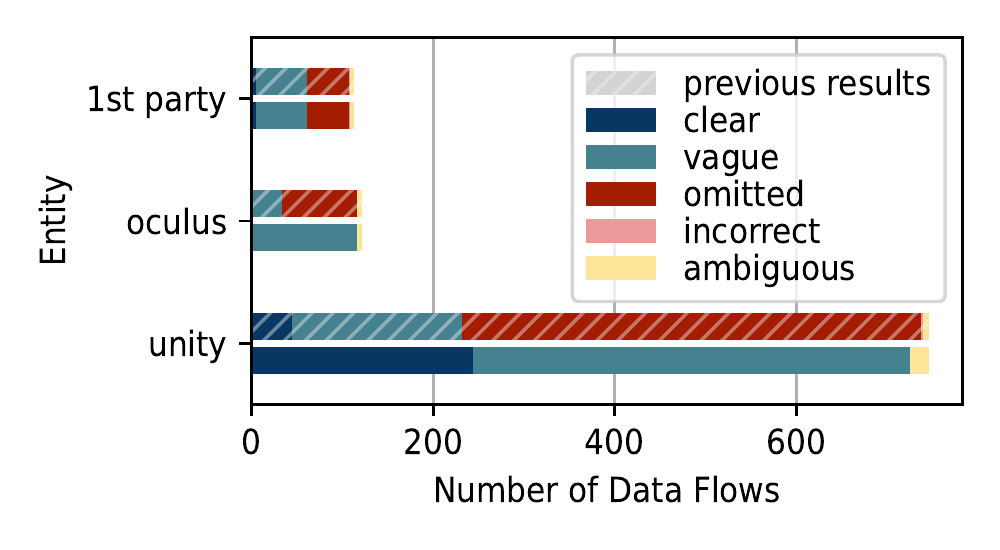}
	\caption{\textbf{Comparisons of \term{network}-to-policy consistency analysis results.} In this graph we compare the results from including Unity and Oculus privacy policies by default and previous results. Previous results are shown by the (rasterized) top bars.}
	\label{fig:policheck-bar-entity-alt}
\end{figure}
}

\vspace{-5pt}
\myparagraph{Data type consistency.} 
\figref{}~\ref{fig:policheck-bar-dtype-5types} reports network-to-policy consistency analysis results by data types---recall that in \secref{}~\ref{sec:labeling-data-flow} we introduced all the exposed data types into three categories: \textit{PII}, \textit{Fingerprint}, and \textit{VR Sensory Data}.
The PII category contributes to 22\% (250/1,135) of all data flows.
Among the three categories, PII has the best consistency: 57\% (142/250) data flows in this category are classified as consistent disclosures.
These data types are well understood and also treated as PII in other platforms. 
On Android~\cite{andow2020actions}, it is reported that 59\% of PII flows were consistent---this is similar to our observation on \oculusvr{}.
The Fingerprint category constitutes 69\% (784/1,135) of all data flows: around 25\% (199/784) of data flows in this category are classified as consistent disclosures. 
The VR Sensory Data category constitutes around 9\% (101/1,135) of all data flows---this category is unique to the VR platform. Only 18\% (18/101) data flows of this category are consistent---this indicates that the collection of data types in this category is not properly disclosed in privacy policies.
\myparagraph{Entity consistency.}
\figref{}~\ref{fig:policheck-bar-entity-5types} reports our network-to-policy consistency results, by entities.
Only 29\% (298/1,022) of third-party and platform data flows are classified as consistent disclosures.
First-party data flows constitute 10\% (113/1,135) of all data flows: 54\% (61/113) of these first-party data flows are classified as consistent disclosures.
\minorrevision{
Thus, 69\% (785/1,135) of all data flows are classified as inconsistent disclosures.} 
Third-party and platform data flows constitute 90\% (1,022/1,135) of all data flows---surprisingly, only 29\% (298/1,022) of these third-party and platform data flows are classified as consistent disclosures.

Unity is the most popular third-party entity, with   66\% (746/1,135) of all data flows. Only 31\% (232/746) of these Unity data flows are classified as consistent, while the majority (69\%) are classified as inconsistent disclosures. 
Platform (\ie Oculus and Facebook) data flows account for 11\% (122/1,135) of all data flows; only 28\% (34/122) of them are classified as consistent disclosures. 
Other less prevalent entities account only around 14\% (154/1,135) of all data flows.

\myparagraph{Referencing Oculus and Unity privacy policies.}
Privacy policies can link to each other. For instance, when using \device{}, users should be expected to consent 
to the Oculus privacy policy (for \oculusvr{}). Likewise, when app developers utilize a third party engine (\eg Unity) their privacy policies should  include the Unity privacy policy. To the best of our knowledge, this aspect has not been considered in prior work~\cite{zimmeck2017automated,andow2020actions,lentzsch2021heyalexa}.

Interestingly, when we included the Oculus and Unity privacy policies (when applicable) in addition to the app's own privacy policy, we found that the majority of platform (116/122 or 96\%) and Unity (725/746 or 97\%) data flows get classified as consistent disclosures. 
\minorrevision{
Thus, 74\% (841/1,135) of all data flows get classified as consistent disclosures.}
%
%
\figref{}~\ref{fig:policheck-bar-entity-alt} shows the comparison of the results from this new experiment  with the previous results shown in \figref{}~\ref{fig:policheck-bar-entity-5types}. 
 These show that data flows are properly disclosed in Unity and Oculus privacy policies even though the app developers' privacy policies usually do not refer to these two important privacy policies.
Furthermore, we noticed that the Oculus and Unity privacy policies are well-written and clearly disclose collected data types. 
As discussed in \cite{andow2020actions}, developers may be unaware of their responsibility to disclose third-party data collections, or they may not know exactly how third-party SDKs in their apps collect data from users. This is a recommendation for future improvement.


\begin{figure}[t]
	\centering
	\includegraphics[width=0.85\columnwidth]{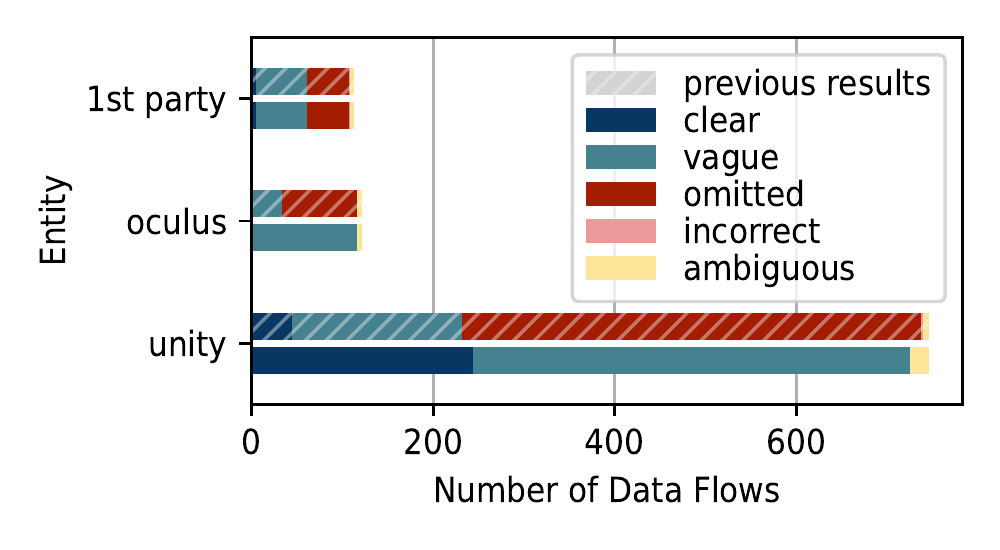}
	\vspace{-15pt}
	\caption{\textbf{Referencing Oculus and Unity privacy policies.} Comparing the results from the ideal case (including Unity and Oculus privacy policies by default) and the previous actual results (only including the app's privacy policy and any third-party privacy policies linked explicitly therein).}
	\label{fig:policheck-bar-entity-alt}
     \vspace{-5pt} 
\end{figure}

\myparagraph{\minorrevision{Validation of PoliCheck results (network-to-policy consistency).}}\minorrevision{To test the correctness of PoliCheck when applied to VR apps, we \minorrevision{manually inspected all data flows from apps that provided a privacy policy,} and  checked their consistency with corresponding collection statements in the policy. Three authors had to agree on the consistency result (one of the five disclosure types) of each data flow. We found the following.}

\minorrevision{
First, we considered {\em multi-class classification} into consistent, omitted and incorrect disclosures, similar to \policheck{}'s evaluation \cite{andow2020actions}. The performance of multi-class classification can be assessed using micro-averaging or macro-averaging of metrics across classes. {\em Micro-averaging} is more appropriate for imbalanced datasets and was also used for consistency analysis of Android apps~\cite{andow2020actions} and Alexa skills~\cite{lentzsch2021heyalexa}. In our VR dataset, we obtained 84\% micro-averaged precision, recall and F1-score}\footnote{
\minorrevision{In multi-class classification, every misclassification is a false positive for one class and a false negative for other classes; thus, micro-averaged precision, recall, and F1-score are all the same (see \appdx{\appref{}~\ref{app:policheck-validation}}).
}}.
\minorrevision{This is comparable to the corresponding numbers when applying PoliCheck to mobile \cite{andow2020actions} and Alexa Skills \cite{lentzsch2021heyalexa}, which reported 90.8\% and 83.3\% (micro-averaged) precision/recall/F1-score, respectively.  For completeness, we also computed the {\em macro-averaged} precision, recall and F1-score to be 74\%, 89\%, and 81\% respectively (\appdx{see~\tabref{}~\ref{tab:policheck-eval}}).
}

\minorrevision{
Second, we considered the binary classification case (\ie{} we treat inconsistent disclosures as positive and consistent disclosures as negative samples). We obtained 77\% precision, 94\% recall, and 85\% F1-score (see \appdx{\appref{}~\ref{app:policheck-validation}} for more details). 
Overall, \policheck{}, along with our improvements for \tool{}, works well on VR apps}\footnote{\minorrevision{However, the precision is lower when distinguishing between clear and vague disclosures. Our validation shows 23\% vague disclosures were actually clearly disclosed. This is because \tool{}'s privacy policy analyzer inherits the limitations of \policheck's NLP model which cannot extract data types and entities from a collection statement that spans multiple sentences.}.}.

\begin{figure*}[t!]
	\centering
	\vspace{-10pt}
	\includegraphics[width=0.9\linewidth]{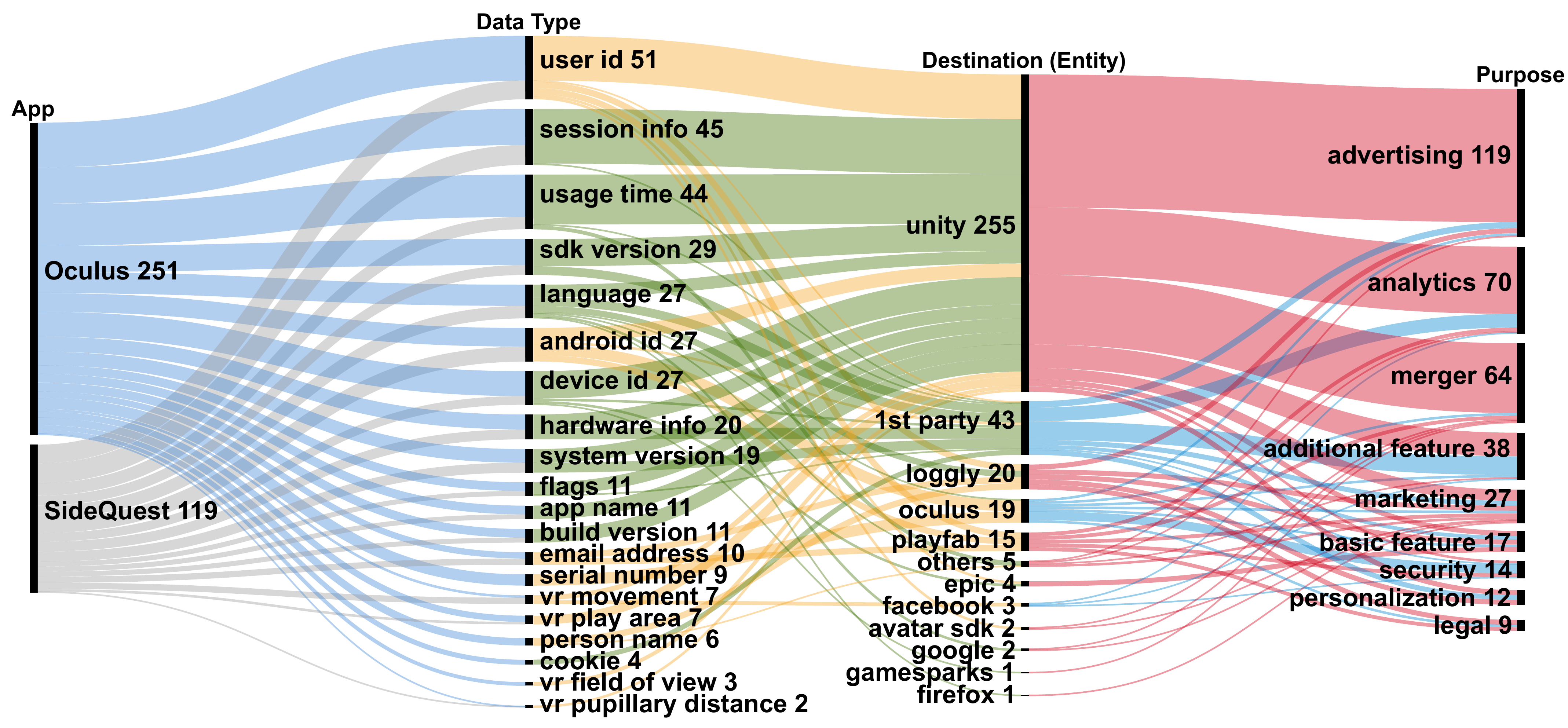}
	\caption{\textbf{Data flows in context.} We consider the data flows (\dataflowtuple) found in the network traffic, and, in particular, the 370 data flows associated with consistent disclosures. 
	We analyze these in conjunction with their {\em purpose} as extracted from the privacy policy text \minorrevision{and depict the augmented tuples {\textit{$\langle$app, data type, destination, purpose $\rangle$}} in the above alluvial diagram. The diagram is read from left to right, for example: (1) out of 251 data flows from the Oculus app store, no more than 51 data flows collect User ID and send it to various destinations; (2) the majority of User ID is collected by Unity; and (3) Unity is responsible for the majority of data flows with the purpose of advertising. Finally, the color scheme of the edges helps keep track of the flow. From App to Data Type, the color indicates the app store: blue for Oculus apps and gray for SideQuest apps. From Data Type to Destination, the color indicates the type of data collected: PII and VR Sensory Data data flows are in orange, while Fingerprinting data flows are in green. From Destination to Purpose, we use blue to denote first-party destinations and red to denote third-party destinations.} 
	}
	\label{fig:polisis-collected}
	\vspace{-10pt}
\end{figure*}


\subsection{Data Collection in Context}
\label{sec:polisis-context}

Consistent (\ie clear, or even vague) disclosures are desirable because they notify the user about the VR apps' data collection and sharing practices. However, they are not sufficient to determine whether the information flow is within its context or social norms.  This context includes (but is not limited to) the purpose and use, notice and consent, whether it is legally required, and other aspects of the ``transmission principle'' in the terminology of contextual integrity~\cite{privacy-in-context}. In the previous section, we have discussed the {\em consistency} of the network traffic \wrt the privacy policy statements: this provides some context. In this section, we identify an additional context: we focus on the {\em purpose} of data collection.

\myparagraph{Purpose.} 
We extract purpose from the app's privacy policy using \polisys{}~\cite{harkous2018polisis}---an online privacy policy analysis service based on deep learning. \polisys{} annotates privacy policy texts with purposes at text-segment level. We developed a translation layer to map annotated purposes from \polisys{} into consistent data flows (see \appdx{\appref{}~\ref{app:polisis-integration}}). This mapping is possible only for data flows with consistent disclosures, since we need the policy to extract the purpose of a data flow.
\minorrevision{We were able to process 293 (out of 359) consistent data flows}\footnote{
\minorrevision{
\polisys{} did not process the text segments that correspond to the remaining 66 consistent data flows: it did not annotate the text segments and simply reported that their texts were too short to analyze.}}
\minorrevision{ that correspond to 141 text segments annotated by \polisys{}.
Out of the 293 data flows, 69 correspond to text segments annotated as ``unspecific'', \ie  \polisys{} extracted no purpose. The remaining 224 data flows correspond to text segments annotated with purposes.
Since a data flow can be associated with multiple purposes, we expanded \minorrevision{the 224} into 370 data flows, so that each data flow has exactly one purpose.} There are nine distinct purposes identified by \polisys{} (including advertising, analytics,  personalization, legal, \etc; see \figref{}~\ref{fig:polisis-collected}).


\minorrevision{To further understand whether data collection is essential to app functionality, we distinguish between purposes that  support core functionality (\ie{}  basic features, security, personalization, legal purposes, and merger) and those unrelated to core functionality (\ie{} advertising, analytics, marketing, and additional features)~\cite{pii-nist-report}. 
Intuitively, core functionality indicates services that users expect from an app, such as reading articles from a news app or making a purchase with a shopping app.}
We found that only 31\% (116/370) of all data flows are related to core functionality, while 69\% (254/370) are unrelated. 
Interestingly, 81\% (94/116) of core-functionality-related data flows are associated with third-party entities,  indicating that app developers use cloud services.
\minorrevision{On the other hand, data collection purposes unrelated to core functionality can be used for marketing emails or cross-platform targeted advertisements.}
This is partly also corroborated by our ATS findings in \secref{}~\ref{sec:ats-ecosystem}: 
83\% (211/254) are associated with third-party tracking entities. In OVR, data types can be collected for tracking purposes and used for ads on other mediums (\eg Facebook website) and not on the \device{} device itself.
%

Next, we looked into the data types exposed for different purposes. 
The majority of data flows related to core functionality (56\% or 65/116) expose PII data types, while Fingerprinting data types appear in most (66\% or 173/254) data flows unrelated to functionality. 
We found that 15 data types are collected for functionality: these are comprised of Fingerprinting (41\% or 48/116 data flows) and VR Sensory Data (3\% or 3/116 data flows). We found that 19 data types are collected for purposes unrelated to functionality: these are comprised of PII (26\% or 65/254 data flows) and VR Sensory Data (6\% or 16/254 data flows). Interestingly, VR Movement, VR Play Area, and VR Field of View are mainly used for ``advertising'', while VR Movement and VR Pupillary Distance are used for ``basic features'', ``security'', and ``merger'' purposes~\cite{harkous2018polisis}.

\minorrevision{
\myparagraph{Validation of \polisys{} results (purpose extraction).} In order to validate the results pertaining to purpose extraction, \minorrevision{we read all the 141 text segments previously annotated by \polisys{}. Then, we manually annotated each text segment with one or more purposes (based on the nine distinct purposes identified by \polisys{}). We had three authors look at each segment independently and then agree upon its annotation. We then compared our annotation with the purpose output by \polisys{} for the same segment. We found that this purpose extraction has 80\%, 79\%, and 78\% micro-averaged precision, recall, and F1-score respectively\footnote{Please note that this is multi-label classification. Thus, unlike multi-class classification for \policheck{}, precision, recall, and F1-score are different.}. These micro-averaged results are directly comparable to the \polisys{}' results in \cite{harkous2018polisis}}:
\minorrevision{\tool{}'s purpose extraction works well on VR apps.}
For completeness, we also computed the macro-averaged precision, recall, and F1-score: 81\%, 78\%, and 78\%, respectively. \tabref{}~\ref{tab:polisis-eval} \appdx{in~\appref{}~\ref{app:polisis-integration}} reports the precision, recall, and F1-score for each purpose classification, and their micro- and macro-averages.} 
\minorrevision{
\section{Discussion}
\label{sec:discussion}
}



\minorrevision{
\subsection{VR-Specific Considerations}
\label{sec:major-findings}
VR tracking has unique aspects and trends compared to other ecosystems, including but not limited to the following.
}

\minorrevision{
\myparagraph{VR ads.} The VR advertising ecosystem is currently at its infancy. Our analysis of destinations from the network traffic revealed that ad-related activity was missing entirely from \oculusvr{} at the time of our experiments. Facebook recently started testing on-device ads for Oculus in June 2021 ~\cite{oculus-ads}.
Ads on VR platforms will be immersive experiences instead of flat visual images; for example, Unity uses separate virtual rooms for ads~\cite{unity-virtual-room-ads}. We expect that tracking will further expand once advertising comes into VR (\eg to include tracking how users interact and behave within the virtual ad space). 
As VR advertising and tracking evolve, our \tool{} methodology, system, and datasets will continue to enable analysis that was not previously possible on any VR platforms.}

\minorrevision{
\myparagraph{Comparison to other ecosystems.} Our analysis showed that the major players in the
\oculusvr{} tracking ecosystem are currently Facebook and Unity (see \figref{}~\ref{fig:top-esld-ats} and~\ref{fig:flow-to-policy-consistency-analysis}). The more established ecosystems such as mobile and Smart TVs are dominated by Alphabet ~\cite{kollnig2021ios, varmarken2020tv}; they also have a more diverse playing field of trackers (\eg{} Amazon, Comscore Inc., and Adobe)---spanning hundreds of tracking destinations~\cite{kollnig2021ios, varmarken2020tv, shuba2020nomoats}. \oculusvr{} currently has only a few players (\eg Facebook, Unity, Epic, and Alphabet). \tool{} can be a useful tool for continuing the study on this developing ecosystem. 
}


\minorrevision{
\myparagraph{Sensitive data.} 
Compared to other devices, such as mobile, Smart TVs and IoT, the type of data that can be collected from a VR headset is arguably more sensitive. For example, \oculusvr{} has access to various biometric information (\eg pupillary distance, hand geometry, and body motion tracking data) that can be used to identify users and even infer their health~\cite{oculus-move}. A study by Miller \etal~\cite{oculus-privacy-3} revealed the feasibility of identifying users with a simple machine learning model using less than five minutes of body motion tracking data from a VR device. Our experiments found evidence of apps collecting data types that are unique to VR, including biometric-related data types (see \secref{}~\ref{sec:labeling-data-flow}). While the numbers we found are small so far, with the developing VR tracking ecosystem, it is important to have a system such as \tool{} to detect the increasing collection of sensitive data over time.
}

\minorrevision{
\myparagraph{Generalization.}
Within \oculusvr{}, we only used \tool{} to analyze 140 apps in our corpus. However, we believe that it can be applied  to other \oculusvr{} apps, as long as they are created according to \oculusvr{} standards.
Beyond \oculusvr{}, the network traffic analysis and network-to-policy consistency analysis can also be applied to other platforms, as long as their network traffic can be decrypted, as was the case with prior work on Android, Smart TV, \etc~\cite{shuba2016antmonitor,razaghpanah2018apps,mohajeri2019watching,varmarken2020tv}.}

\minorrevision{
\subsection{Recommendations}
\label{sec:recommendations}
Based on our findings, we provide recommendations for the \oculusvr{} platform and developers to improve their data transparency practices.}

\minorrevision{
\myparagraph{Provide a privacy policy.}
We found that 38 out of the 140  popular apps, out of which 19 are from the Oculus app store,  did not provide any privacy policy at all.
Furthermore, 97\% of inconsistent data flow disclosures were due to omitted disclosures by these 38 apps missing privacy policies (see \secref{}~\ref{sec:privacy-policy}).
We recommend that the \oculusvr{} platform require developers to provide a privacy policy for their apps, especially those available on the official Oculus app store.
}

\minorrevision{
\myparagraph{Reference other parties' privacy policies.} 
Developers are not the only ones collecting data during the usage of an app: third-parties (\eg{} Unity, Microsoft) and platform-party (\eg{} Oculus/Facebook) can also collect data.  We found that 81 out of 102 app privacy policies did not reference policies of third-party  libraries used by the app. We recommend that developers reference third-party and platform-party privacy policies. If they do that, then the consistency of disclosures will be quite high: up to 74\% of data flows in the network traffic we collected (see~\secref{}~\ref{sec:policheck-results}). This indicates that, at least at this early stage, the VR ecosystem is better behaved than the mobile tracking ecosystem. 
}

\minorrevision{
\myparagraph{Notice and consent.}
We found that fewer than 10 out of 102 apps that provide a privacy policy explicitly ask users to read it and give consent to data collection (\eg{} for analytics purposes) upon first opening the app.
We recommend that developers provide notice and ask for users' consent (\eg when a user launches the app for the first time) for data collection and sharing, as required by privacy laws such as GDPR~\cite{gdpr-implementation}.}

\minorrevision{
\myparagraph{Notifying developers.} We contacted Oculus as well as the developers of the 140 apps that we tested. We provided courtesy notifications of the specific data flows and consistency we identified in their apps, along with recommendations. We received 24 responses (see the details \appdx{in~\appref{}~\ref{sec:responses}}).}
\minorrevision{Developers were, in general, appreciative of the information and willing to adopt recommendations to improve their privacy policies. Several indicated they did not have the training or tools to ensure consistent disclosures.}

\section{Related Work}
\label{sec:related-work}

\paragraph{Privacy in Context.}
The framework of ``Privacy in Context"~\cite{privacy-in-context} specifies the following aspects of information flow: (1) actors: sender, recipient, subject; (2) type of information; and (3) transmission principle. The goal is to determine whether the information flow is appropriate within its context. The ``transmission principle" is key in determining the appropriateness of the flow and may  include: the purpose of data collection, notice and consent, required by law, \etc~\cite{privacy-in-context}. In this paper, we seek to provide context for the data flows (\dataflowtuple{}) found in the network traffic. We primarily focus on the network-to-policy {\em consistency}, {\em purpose} of data collection, and we briefly comment on notice and consent. Most prior work on network analysis only characterized destinations (first \vs third parties, ATS, \etc) or data types exposed without additional contexts. One exception is MobiPurpose~\cite{jin2018mobipurpose}, which  inferred data collection purposes of mobile (not VR) apps, using network traffic and app features (\eg URL paths, app metadata, domain name, \etc); the authors stated that ``the purpose interpretation can be subjective and ambiguous''. Our notion of purpose is explicitly stated in the privacy policies and/or indicated by the destination domain matching ATS blocklists.
Shvartzshnaider \etal introduced the contextual integrity (CI) framework to understand and evaluate privacy policies~\cite{shvartzshnaider2019going}---they, however, leveraged manual inspection and not automation.


\myparagraph{\minorrevision{Privacy of various platforms.}}
The research community has looked into privacy risks in various platforms, using static or dynamic code analysis, and---most relevant to us---network traffic analysis.
Enck \etal performed static analysis of Android apps~\cite{androidappstudy} and discovered PII misuse (\eg personal/phone identifiers) and ATS activity.
Taintdroid, first introduced taint tracking for mobile apps~\cite{taintdroid}.
Ren \etal~\cite{ren2019information} did a comprehensive evaluation of information exposure on smart home IoT devices. 
Moghaddam \etal and Varmarken \etal observed the prevalence of PII exposures and ATS activity~\cite{mohajeri2019watching,varmarken2020tv} in Smart TVs.
Lentzsch \etal~\cite{lentzsch2021heyalexa} performed a comprehensive evaluation on Alexa, a voice assistant platform.
Ren \etal~\cite{recon}, 
Razaghpanah \etal~\cite{razaghpanah2018apps}, and Shuba \etal~\cite{shuba2016antmonitor,shuba2018nomoads,shuba2020nomoats} developed tools for analysis of network traffic generated by mobile apps, and inspection for privacy exposures and ATS activity.
Our work is the first to perform network traffic analysis on the emerging \oculusvr{} platform, using dynamic analysis to capture and decrypt networking traffic on the device; this is more challenging for Unity and Unreal based apps because, unlike prior work that dealt with standard Android APIs, we had to deal with stripped binary files (\ie no symbol table). 
\minorrevision{
Augmented reality (AR) is another platform the research community has been focusing on in the past decade~\cite{roesner2014security,acquisti2014face,wassom2014augmented,kotsios2015privacy,rauschnabel2018antecedents,lebeck2018towards}.
While AR augments our perception and interaction with the real world, VR replaces the real world with a virtual one.
Nevertheless, some AR privacy issues are similar to those in VR since they have similar sensors, \eg motion sensors. }

\mycomment{
\rahmadi{
\descr{Privacy of AR.}
\todo{Maybe just change this into just one sentence or something like that.}
Augmented reality (AR) is another technology, orthogonal to VR, that the research community has been looking into in the past decade.
Roesner~\etal outlined the security and privacy concerns of AR systems~\cite{roesner2014security}.
Acquisti~\etal investigated the implications of the increasing public availability of facial images online and the rise of computer recognition capability in the context of AR 
applications~\cite{acquisti2014face}.
Wassom~\etal explored the law, privacy, and ethical aspects of AR systems~\cite{wassom2014augmented}.
Other work studied the privacy risks of AR smart glasses~\cite{kotsios2015privacy,rauschnabel2018antecedents} and other AR devices~\cite{lebeck2018towards} from user perspectives. 
While AR augments our perception and interaction with the real world, VR replaces the real world with a virtual one.
Nevertheless, some of AR privacy issues are similar to those in VR since they have similar sensors, \eg motion sensors. 
}
}

\descr{Privacy of VR.}
Although there is work on security aspects of VR devices (\eg{} authentication and attacks on using virtual keyboards)  \cite{luooculock,8802674,duezguen2020towards,mathis2021fast},
%
%
the privacy of VR is currently not fully understood.
Adams \etal~\cite{adams2018ethics} interviewed VR users and developers on security and privacy concerns, and learnt that they were concerned with data collection potentially performed by VR devices (\eg{} sensors, device being always on) and that they did not trust VR manufacturers (\eg{} Facebook owning Oculus).
\minorrevision{Miller \etal present a study on the implications of the ability of VR technology to track body motions~\cite{oculus-privacy-3}.}
%
Our work is motivated by these concerns but goes beyond user surveys to analyze data collection practices exhibited in the network traffic and stated in privacy policies. 

\descr{Privacy policy analysis.}
Privacy policy and consistency analysis in various app ecosystems ~\cite{zimmeck2014privee,slavin2016toward,zimmeck2017automated,wang2018guileak,harkous2018polisis,andow2019policylint, andow2020actions} is becoming increasingly automated.
Privee~\cite{zimmeck2014privee} is a privacy policy analyzer that uses NLP  to classify the content of a website privacy policy using a set of binary questions.
Slavin \etal used static code analysis, ontologies, and information flow analysis to analyze privacy policies for mobile apps on Android~\cite{slavin2016toward}.
Wang \etal applied similar techniques to check for privacy leaks from user-entered data in GUI~\cite{wang2018guileak}.
Zimmeck \etal also leveraged static code analysis for privacy policy consistency analysis~\cite{zimmeck2017automated}; they improved on previous work by attempting to comply with legal requirements (\eg first \vs third party, negative policy statements, \etc).
In \secref{}~\ref{sec:privacy-policy}, we leverage two state-of-the-art tools, namely  \policheck{}~\cite{andow2020actions} and \polisys{}~\cite{harkous2018polisis}, to perform data-flow-to-policy consistency analysis and to extract the data collection purpose, respectively.
\policheck{} was built on top of \policylint{}~\cite{andow2019policylint}, a privacy policy analyzer for mobile apps. It analyzes both positive and negative data collection (and sharing) statements, and detects contradictions.
Lentzsch \etal also used off-the-shelf \policheck{}  
 using a data ontology crafted for Alexa skills.
\tool{} focuses on \oculusvr{} and improves on \policheck{} in several ways, including VR-specific ontologies, referencing third-party policies, and extracting data collection purposes.

\minorrevision{
\section{Conclusion}
\label{sec:limitation-conclusion}
}

\minorrevision{
\paragraph{Summary.}
We present the first comprehensive study of privacy aspects for Oculus VR (\oculusvr{}), the most popular VR platform. 
We developed \tool{}, a methodology and system to characterize the data collection and sharing practices of the \oculusvr{} ecosystem by (1) capturing and analyzing data flows found in the network traffic of 140 popular \oculusvr{} apps, and (2) providing additional contexts via privacy policy analysis that checks for consistency and identifies the purpose of data collection.
\minorrevision{We make \tool{}'s implementation and datasets publicly available at \cite{ovrseen-release}.}} \inusenixversion{An extended version of this paper, including appendices, can be found in~\cite{trimananda2021auditing}.}\inarxivversion{This is the extended version of our paper, with the same title, published at USENIX Security Symposium 2022. Please take a look at our project page for more information~\cite{ovrseen-release}.}

\myparagraph{Limitations and future directions.}  
On the networking side, we were able to decrypt, for the first time, traffic of \oculusvr{} apps, but the \oculusvr{} platform itself is closed  and we could not decrypt most of its traffic. In future work, we will explore the possibility of addressing this limitation by further exploring binary analysis. 
On the privacy policy side, \policheck{} and \polisys{} rely on different underlying NLP model, with inherent limitations and incompatibilities---this motivates future work on a unified privacy policy and context analyzer.

\minorrevision{
\section*{Acknowledgment}
\label{sect:acknowledgment}

This project was supported by NSF Awards 1815666 and 1956393.
We would like to thank our shepherd, Tara Whalen, and the USENIX Security 2022 reviewers for their feedback, which helped to significantly improve the paper. 
We would also like to thank Yiyu Qian, for his help with part of our data collection process. 
}

\bibliographystyle{abbrv}
{
\bibliography{master,online}
}

\newpage
\appendix
\ifthenelse{\value{isusenixversion}>0}
{

\section{Data Privacy on Oculus}
\label{sec:oculusdataprivacy}

\section{Network Traffic Collection Details}

\label{app:net-traffic-appendix}
\begin{figure}[tb!]
	\centering
	\includegraphics[width=1\linewidth]{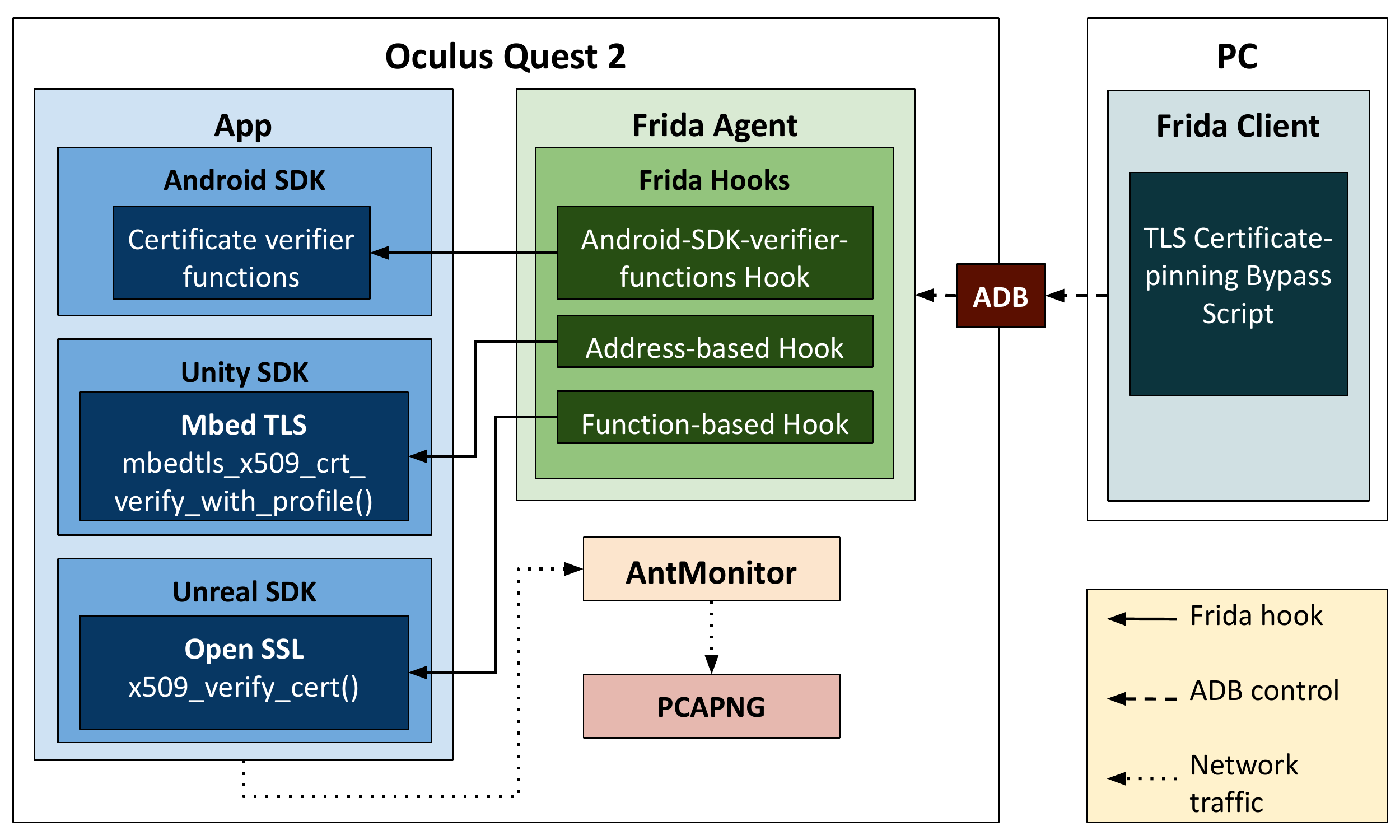}
	\caption{Network traffic collection and decryption.}
	\label{fig:traffic-collection}
\end{figure}

\subsection{Improving AntMonitor}
\label{app:antmonitor-improvement}

\subsection{Binary Analysis Workflow}
\label{app:binary-analysis}
\begin{figure}[tb!]
	\centering
	\includegraphics[width=0.95\linewidth]{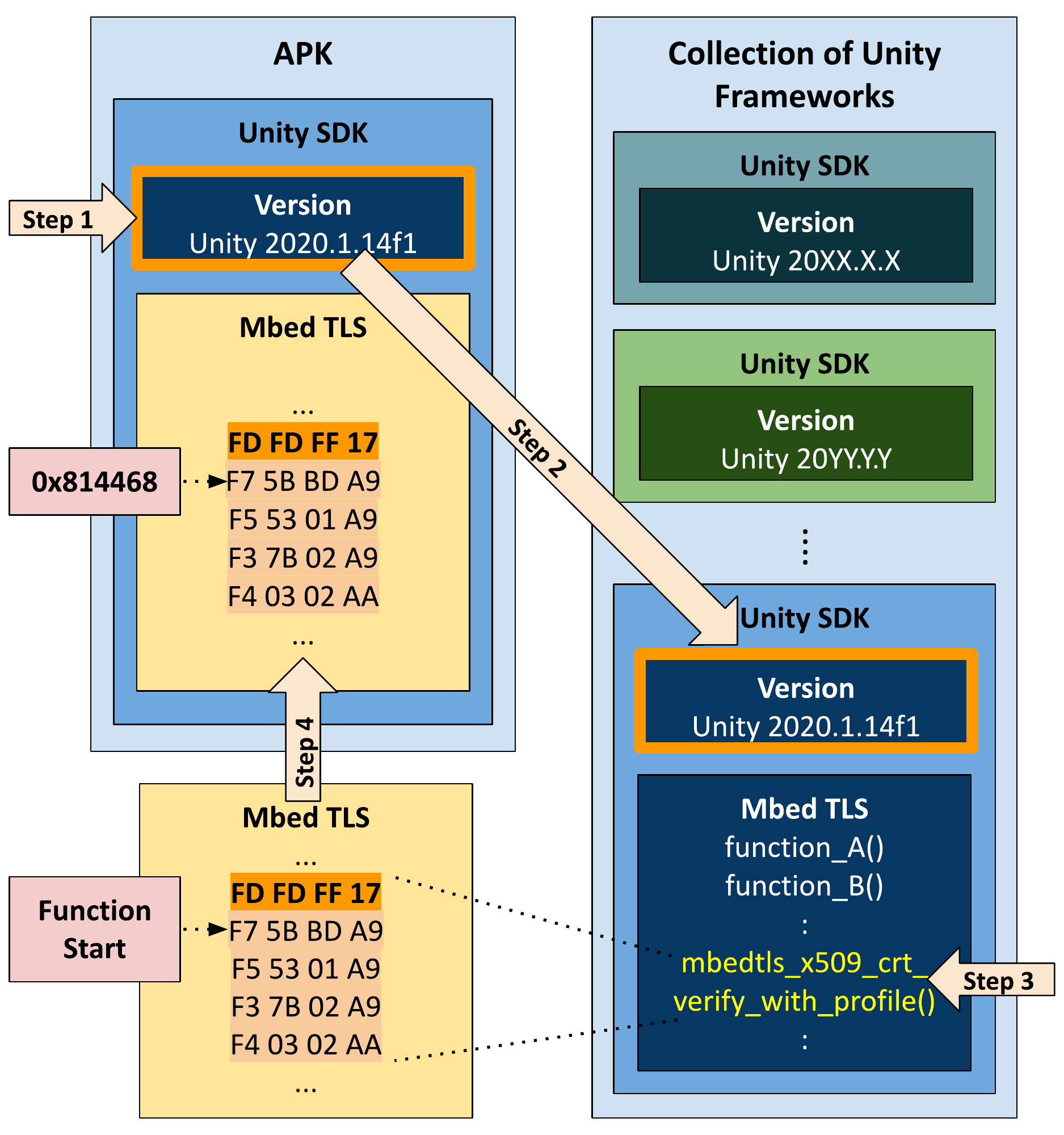}
	\caption{\minorrevision{\textbf{Our decryption technique.}} Example on Spatial, an app that enables people to meet through VR~\cite{spatial-app}.}
	\label{fig:novel-decryption-technique}
\end{figure}

\section{ATS and Data Types Details}
 \label{app:ats-data-types-appendix}

\subsection{Extracting Data Types}
\label{app:extracting-data-types-appendix}

\begin{table*}[htbp]
	\footnotesize
	\centering
	\begin{tabularx}{\linewidth}{ l | p{4cm} | X }
	    \toprule
		\textbf{PII} & \textbf{Finer Grain Data Types} &  \textbf{Keywords or ~\texttt{Regular Expressions}} 
		 \\
		\midrule
        Android ID & - &
          (hard-coded Android ID), android\_id, x--android--id 
        \\ \midrule
        Device ID & - &
          (hard-coded Oculus Device ID), deviceid, device\_id, device--id
        \\ \midrule
        Email & - & (hard-coded user email address), email  \\ \midrule
	    Geolocation & Country Code, Time Zone, GPS & countryCode, timeZoneOffset, gps\_enabled  \\ \midrule
		Person Name & First Name, Last Name & (hard-coded from Facebook Account)  \\ \midrule
		Serial Number & - & (hard-coded Oculus Serial Number), x--oc--selected--headset--serial \\ \midrule
		User ID & User ID and PlayFab ID & user\_id, UserID, x--player, x--playeruid, profileId, anonymousId, PlayFabIDs \\
		\midrule	\textbf{Fingerprint} \\ 
		\midrule
        App Name & App Name and App Version & app\_name, appid, application\_name, applicationId, X--APPID, gameId, package\_name, app\_build, localprojectid, android\_app\_signature, gameVersion, package\_version \\ \midrule
        Build Version & - & build\_guid, build\_tags  \\ \midrule 
        Cookies & - & cookie \\ \midrule
        Flags & Do Not Track, Tracking, Jail Break, Subtitle On, Connection Type, Install Mode, Install Store, Scripting Backend  &  x--do--not--track, tracking, rooted\_or\_jailbroken, rooted\_jailbroken, subtitles, connection--type, install\_mode, install\_store, device\_info\_flags, scripting\_backend \\ \midrule
        Hardware Info & Device Model, Device RAM, Device VRAM, CPU Vendor, CPU Flags, Platform CPU Count and Frequency, GPU Name,  GPU Driver, GPU Information, OpenGL Version, Screen Resolution, Screen DPI, Fullscreen Mode, Screen Orientation, Refresh Rate, Device Info, Platform & 
        device\_model, device\_type, enabled\_vr\_devices, vr\_device\_name, vr\_device\_model,
        Oculus/Quest/hollywood, \verb "Oculus[+ ]?Quest", \verb "Quest[ ]?2", device\_ram, device\_vram, Qualcomm Technologies, Inc KONA, ARM64 FP ASIMD AES, ARMv7 VFPv3 NEON, cpu\_count, cpu\_freq,
        ARM64+FP+ASIMD+AES, arm64-v8a,+armeabi-v7a,+armeabi, Adreno (TM) 650, GIT@09c6a36, GIT@a8017ed,
        gpu\_api, gpu\_caps, gpu\_copy\_texture\_support, gpu\_device\_id, gpu\_vendor\_id, gpu\_driver, gpu\_max\_cubemap\_size, gpu\_max\_texture\_size, gpu\_shader\_caps, gpu\_supported\_render\_target\_count, gpu\_texture\_format\_support, gpu\_vendor, gpu\_version, OpenGL ES 3.2, \verb "\+3664,\+1920" , 3664 x 1920, 3664x1920, width=3664, screen\_size, screen\_dpi, is\_fullscreen, screen\_orientation, refresh\_rate,
        device\_info\_flags, releasePlatform, platform, platformid
        \\ \midrule
        SDK Version & Unity Version, Unreal Version, Client Library, VR Shell Version & 
        \verb "Unity[- ]?v?20[12]\d\.\d+\.\d+", \verb "Unity[- ]?v?[0-6]\.\d+\.\d+", \verb "UnityPlayer/[\d.]+\d", \verb "UnitySDK-[\d.]+\d", x--unity--version, sdk\_ver, engine\_version, X--Unity--Version, sdk\_ver\_full, ARCore,
        X-UnrealEngine-VirtualAgeStats, engine=UE4, UE4 0.0.1
        clientLib, clientLibVersion, x--oc--vrshell--build--name
        \\ \midrule
        Session Info & App Session, Session Counts, Events, Analytics, Play Session Status, Play Session Message, Play Session ID & AppSession, session\_id, sessionid, event-id, event\_id, objective\_id, event-count, event\_count, session-count, session\_count, analytic, joinable, lastSeen, join\_channel, JoinParty, SetPartyActiveGameOrWorldID, SetPartyIDForOculusRoomID, JoinOpenWorld, partyID, worldID, gameOrWorldID, oculusRoomID \\ \midrule
        System Version & - & (hard-coded OS version strings), x-build-version-incremental, os\_version, operatingSystem, os\_family \\ \midrule
        Usage Time & Start Time, Duration & t\_since\_start, startTime, realtimeDuration, seconds\_played, game\_time, gameDuration \\ \midrule
        Language & - & language, language\_region, languageCode, system\_language \\
        
		\midrule	\textbf{VR Sensory Data} \\ 
		\midrule
        VR Field of View & - & vr\_field\_of\_view  \\ \midrule
        VR Movement & Position, Rotation, Sensor Flags & vr\_position, vr\_rotation, gyroscope, accelerometer, magnetometer, proximity, sensor\_flags, left\_handed\_mode  \\ \midrule
        VR Play Area & Play Area, Play Area Geometry, Play Area Dimention, Tracked Area Geometry, Tracked Area Dimension & vr\_play\_area\_geometry, vr\_play\_area\_dimension, playarea, vr\_tracked\_area\_geometry, vr\_tracked\_area\_dimension \\ \midrule
        VR Pupillary Distance & - & vr\_user\_device\_ipd   \\

	    \bottomrule
	\end{tabularx}
	\caption{\textbf{Extracting Data Types.} Summarizes how we group data types and the keywords and regular expressions (italicized) that we use to identify them (\secref{}~\ref{sec:network-traffic-dataset} and \appref{}~\ref{app:extracting-data-types-appendix}).}
	\label{tab:data-type-extraction-appendix}
\end{table*}

\subsection{Missed by Blocklists}
\label{app:missed-blocklists}

\begin{table*}[tb!]
	\footnotesize
	\centering
	\begin{tabularx}{\linewidth}{p{6cm} p{2.5cm} p{8cm} }
	    \toprule
		\textbf{FQDN} & \textbf{Organization} & \textbf{Data Types} 
		\\
		\midrule
		bdb51.playfabapi.com, 1c31b.playfabapi.com & Microsoft & Android ID, User ID, Device ID, Person Name, Email, Geolocation, Hardware Info, System Version, App Name, Session Info, VR Movement  \\
		\midrule
		sharedprod.braincloudservers.com & bitHeads Inc. & User ID, Geolocation, Hardware Info, System Version, SDK Version, App Name, Session Info, Language  \\
		\midrule
		cloud.liveswitch.io & Frozen Mountain Software & User ID, Device ID, Hardware Info,  System Version, App Name, Language, Cookie \\
		\midrule
		datarouter.ol.epicgames.com & Epic Games & User ID, Device ID, Hardware Info, SDK version, App Name, Session Info  \\
		\midrule
		9e0j15elj5.execute-api.us-west-1.amazonaws.com & Amazon & User ID, Hardware Info, System Version, SDK Version, Usage Time  \\ 
		\midrule
		63fdd.playfabapi.com & Microsoft & Android ID, User ID, Email, SDK Version, App Name \\ 	\midrule
		us1-mm.unet.unity3d.com & Unity & Hardware Info, System Version, SDK Version, Usage Time \\ 	\midrule
		scontent.oculuscdn.com & Facebook & Hardware Info, System Version, SDK Version \\ \midrule
		api.avatarsdk.com & Itseez3d & User ID, Hardware Info, SDK Version \\	\midrule
		52.53.43.176 & Amazon &  Hardware Info, System Version, SDK Version \\	\midrule
		kingspraymusic.s3-ap-southeast-2.amazonaws.com,
		
		s3-ap-southeast-2.amazonaws.com & Amazon &  Hardware Info, System Version, SDK Version \\	\midrule
		pserve.sidequestvr.com & SideQuestVR &  Hardware Info, System Version, Language \\	\midrule
		gsp-auw003-se24.gamesparks.net,
		
		gsp-auw003-se26.gamesparks.net,
		
		gsp-auw003-se30.gamesparks.net,
		
		live-t350859c2j0k.ws.gamesparks.net & GameSparks &  Device ID, Flags \\	\midrule
		yurapp-502de.firebaseapp.com & Alphabet &  Hardware Info, SDK Version \\	
	    \bottomrule
	\end{tabularx}
	\caption{\textbf{Missed by Blocklists Continued.} We provide third-party FQDNs that are missed by blocklists based on the number data types that are exposed. This is the full details of Table~\ref{tab:missed-fqdns}.}
	\label{tab:missed-fqdns-appendix}
\end{table*}

\section{Privacy Policy Analysis Details}
\label{app:privacy-policy-analysis-details}

\subsection{Other PoliCheck Improvements}
\label{app:other-policheck-improvements}

\begin{table}[t!]
	\tabletextsize
	\centering

    \minorrevision{
	\begin{tabularx}{\linewidth}{X  r r r r}
	    \toprule
		\textbf{Label} & \textbf{Prec.} & \textbf{Recall} & \textbf{F1} & \textbf{Support}
		\\
		\midrule
		\multicolumn{5}{l}{\textbf{three-class classification}}\\
		\midrule
		consistent     & 0.93 & 0.74 & 0.82 & 454 \\
		incorrect      & 0.50 & 1.00 & 0.67 & 2  \\
		omitted        & 0.77 & 0.94 & 0.85 & 425\vspace{2px}\\
		\textit{macro-average} & 0.74 & 0.89 & 0.81 & \\
		\rowcolor{brandeisblue!10}\textit{micro-average} & 0.84 & 0.84 & 0.84 &  \\
		
		\midrule
		\multicolumn{5}{l}{\textbf{binary classification}} \\
		\midrule
		inconsistent (positive) & 0.77 & 0.94 & 0.85 & 427 \\
		consistent (negative)   & 0.93 & 0.74 & 0.82 & 454 \\
	    \bottomrule
	\end{tabularx}
	}
	\caption{\minorrevision{\textbf{\policheck{} validation.} Multi-class and binary classification metrics for each disclosure type along with the averaged performance. Note that support is in terms of number of data flows.}}
	\label{tab:policheck-eval}
\end{table}


\subsection{PoliCheck Validation}
\label{app:policheck-validation}

\begin{table}[t!]
	\tabletextsize
	\centering

    \minorrevision{
	\begin{tabularx}{\linewidth}{X  r r r r}
	    \toprule
		\textbf{Label} & \textbf{Prec.} & \textbf{Recall} & \textbf{F1} & \textbf{Support}
		\\
		\midrule
		additional service feature     & 0.74 & 0.70 & 0.72 & 20 \\
		advertising                    & 0.94 & 1.00 & 0.97 & 16 \\
		analytics research             & 0.91 & 0.80 & 0.85 & 25 \\
		basic service feature          & 0.82 & 0.45 & 0.58 & 20 \\
		legal requirement              & 0.64 & 1.00 & 0.78 & 9 \\
		marketing                      & 0.92 & 0.75 & 0.83 & 16 \\
		merger acquisition             & 0.78 & 0.88 & 0.82 & 8 \\
		personalization customization  & 0.80 & 0.67 & 0.73 & 6 \\
		service operation and security & 0.82 & 0.64 & 0.72 & 14 \\
		unspecific                     & 0.75 & 0.90 & 0.81 & 49 \\
		\midrule
		\textit{macro-average}        & 0.81 & 0.78 & 0.78 &  \\
		\rowcolor{brandeisblue!10}\textit{micro-average}        & 0.80 & 0.79 & 0.79 &  \\
	    \bottomrule
	\end{tabularx}
	}
	\caption{\minorrevision{\textbf{\polisys{} validation.} Multi-label classification metrics for each purpose along with the averaged performance. Note that support is in terms of number of text segments. Text segments that \polisys{} does not annotate with a purpose is annotated as ``unspecific''.}}
	\label{tab:polisis-eval}
\end{table}

\subsection{Polisis Integration and Validation}
\label{app:polisis-integration}

\mysection{Responses from Developers}
\label{sec:responses}

}
{
{
\centering
\section*{\scshape APPENDICES}
}

\section{Data Privacy on Oculus}
\label{sec:oculusdataprivacy}
In this appendix, we provide more details from our observations on data collection practices on \oculusvr{} that complements our explanation in \secref{}~\ref{sec:background}.
In our preliminary observation, we discovered two findings that motivate us to further study VR privacy in the context of \oculusvr{}.

First, we discovered that Oculus has been actively updating their privacy policy over the years.
We collected different versions of Oculus privacy policy~\cite{oculus-privacy-policy} over time using Wayback Machine 
and examined them manually.
Most notably, we observed a major change in their privacy policy around May 2018. We suspect that this is due to the implementation of the GDPR on May 25, 2018~\cite{gdpr-implementation}---this has required Oculus to be more transparent about its data collection practices.
For example, the privacy policy version before May 2018 declares that Oculus collects information about ``physical movements and dimensions''. 
The version after May 2018 adds ``play area'' as an example for ``physical movements and dimensions''.
Although it has not been strictly categorized as PII, ``play area'' might represent the area of one's home---it loosely contains ``information identifying personally owned property'' that can potentially be linked to an individual~\cite{pii-nist-report}.
This motivates us to empirically study the data collection practices on \oculusvr{}. We report how we use \tool{} to collect network traffic and study data exposures on \oculusvr{} in \secref{}~\ref{sec:network-traffic-analysis}.

\mycomment{
\begin{figure}[!htb]
	\centering
	\includegraphics[width=1\linewidth]{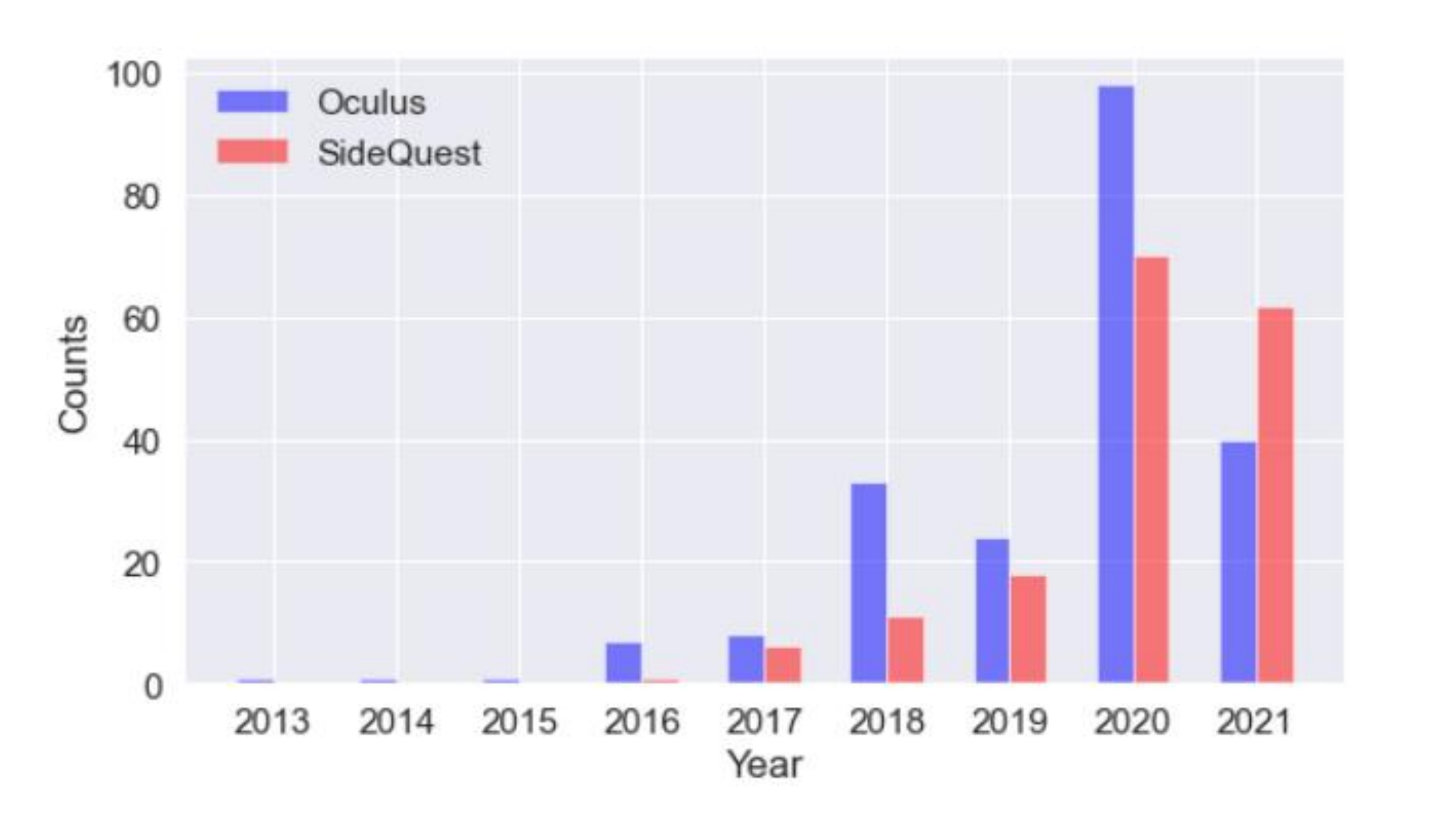}
	\caption{Privacy policy last updated times for apps from the official Oculus and SideQuest app stores.}
	\label{fig:last-updated-times}
\end{figure}
}

Second, we found that many apps do not have privacy policies.
Even if they have one, we found that many developers neglect updating their privacy policies regularly. Many of these privacy policies even do not have \emph{last updated times} information. 
We found that only around 40 (out of 267) apps from the official Oculus app store and 60 (out of 1075) apps from the SideQuest app store have updated their privacy policy texts in 2021.
Thus, we suspect that an app's actual data collection practices might not always be consistent with the app's privacy policy.
This motivates us to study how consistent an app's privacy policy describes the app's actual data collection practices.
We report how we use \tool{} to analyze privacy policies in \secref{}~\ref{sec:privacy-policy}.

\section{Network Traffic Collection Details}
In this appendix, we provide more details on \tool{}'s system for collecting and decrypting network traffic, introduced in~\secref{}~\ref{sec:network-traffic-collection}.
\figref{}~\ref{fig:traffic-collection} depicts a detailed version of the network traffic collection system (\ie \circled{\textbf{1}} in \figref{}~\ref{fig:our-work}), which consists of two main components---AntMonitor and Frida.
In \appref{}~\ref{app:antmonitor-improvement}, we describe the improvements we made to AntMonitor; and in \appref{}~\ref{app:binary-analysis}, we provide the detailed workflow of our automated binary analysis technique for finding addresses of certificate validation functions so that we can hook into them with Frida.

\label{app:net-traffic-appendix}
\begin{figure}[tb!]
	\centering
	\vspace{-5pt}
	\includegraphics[width=1\linewidth]{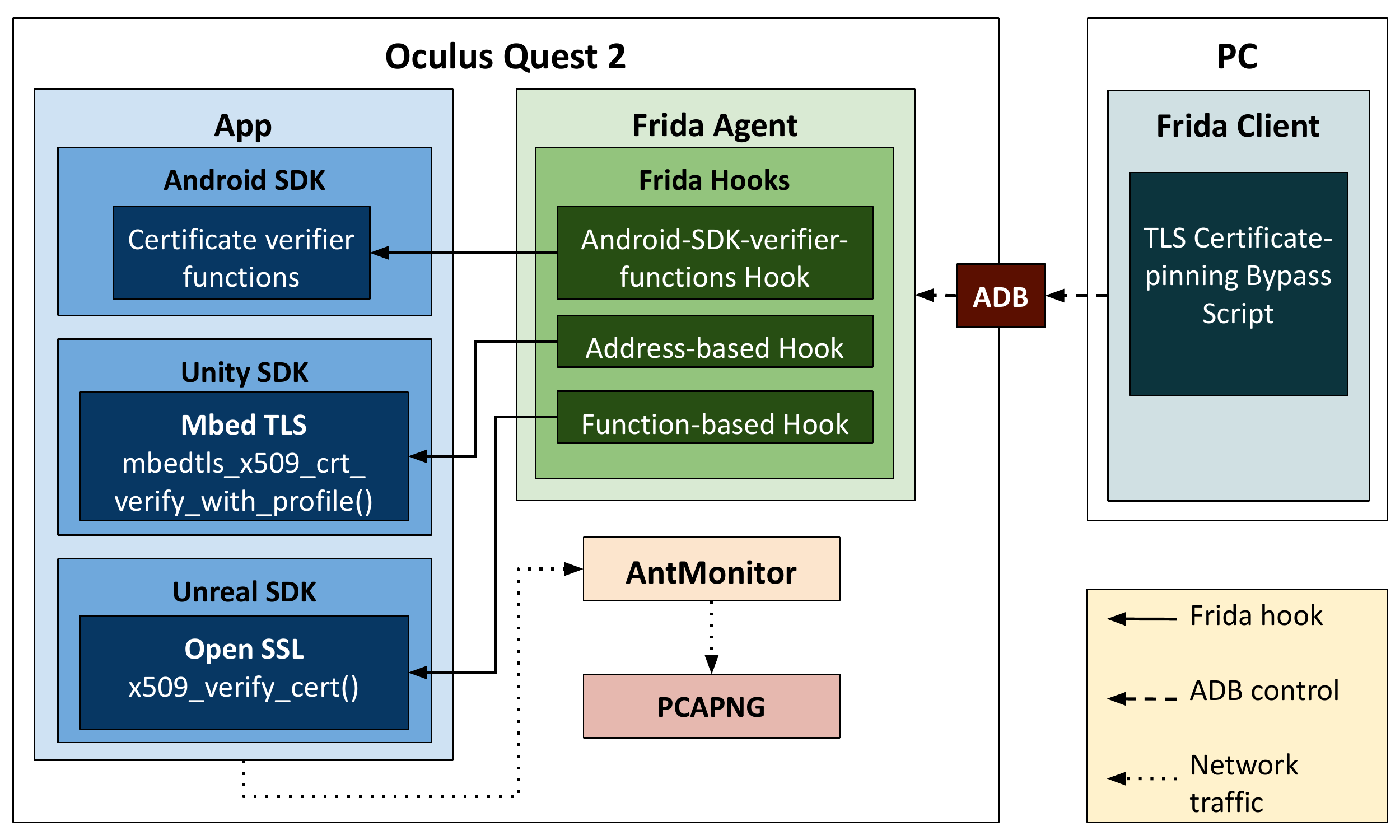}
	\caption{Network traffic collection and decryption.}
	\vspace{-5pt}
	\label{fig:traffic-collection}
\end{figure}

\subsection{Improving AntMonitor}
\label{app:antmonitor-improvement}
The original version of AntMonitor has several limitations, which we address in this paper. First, the released version of AntMonitor supports only up to Android 7. Unfortunately, \device{} runs Oculus OS that is based on Android 10---a version of Android which underwent a multitude of changes to TLS \cite{android10TLS} and filesystem access \cite{android10proc}, effectively breaking AntMonitor's decryption and packet-to-app mapping capabilities. To restore decryption, we first downgraded all connections handled by AntMonitor to TLS 1.2 so that we can extract servers' certificates, which are encrypted in TLS 1.3 (the default TLS version in Android 10). The servers' certificates are needed so that AntMonitor can sign them with its own CA and pass them on to the client app. In addition to downgrading the TLS version, we also updated to new APIs for setting the SNI (server name identification), since the original version used Java reflection to access hidden methods which were no longer available in Android 10. Further, we updated how AntMonitor checks for trusted certificates to remain compatible with Android 10's stricter security requirements. Similarly, in order to fix the packet-to-app mapping, which relied on reading the now-restricted \code{/proc/net} files, we re-implemented the functionality using Android 10's new APIs from the \code{ConnectivityManager}. 

Second, AntMonitor prevents common traffic analysis tools, such as \tshark, from re-assembling TCP streams because it saves both encrypted and decrypted versions of packets in the same PCAPNG file and does not adjust the sequence and ack numbers accordingly. In our work, we wanted to take advantage of \tshark's re-assembly features, namely the desegmentation of TCP streams and HTTP headers and bodies, so that we could analyze parsed HTTP traffic with confidence. To that end, we modified AntMonitor to keep track of decrypted sequence and ack numbers for each decrypted flow and to save decrypted packets in a separate PCAPNG file with their adjusted sequence and ack numbers. 
Without this improvement, the encrypted and decrypted packets would share the same sequence and ack numbers, inhibiting TCP re-assembly.

In order to enable other researchers to continue using AntMonitor in newer Android versions, we will submit a pull request to its open source repository.

\subsection{Binary Analysis Workflow}
\label{app:binary-analysis}
\begin{figure}[tb!]
	\centering
	\includegraphics[width=0.95\linewidth]{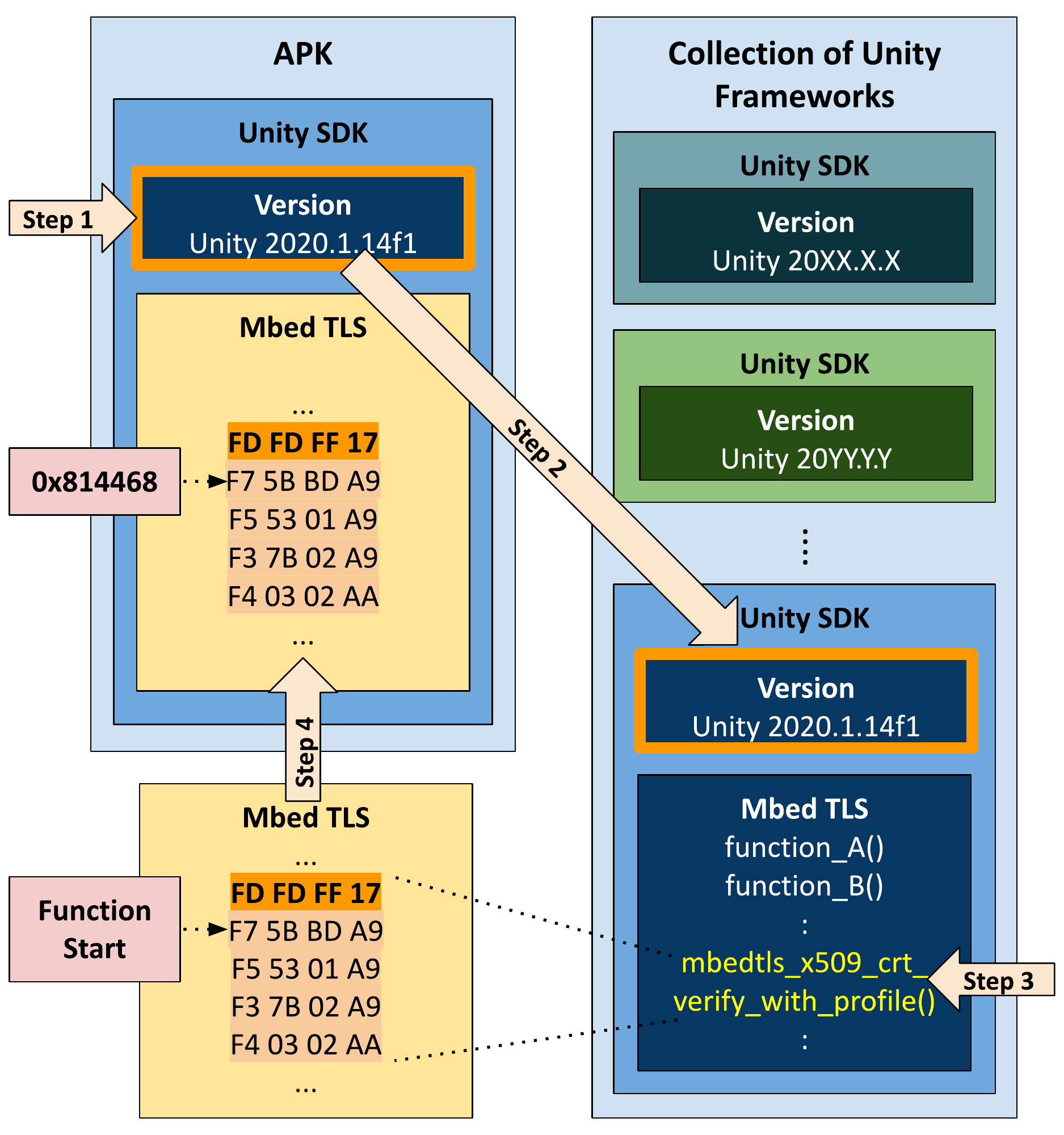}
	\caption{\minorrevision{\textbf{Our decryption technique.}} Example on Spatial, an app that enables people to meet through VR~\cite{spatial-app}.}
	\label{fig:novel-decryption-technique}
\end{figure}

In \secref{}~\ref{sec:network-traffic-collection} we introduced our automated binary analysis technique for finding addresses of certificate validation functions in Unity's ~\cite{unity-mbedtls} library so that we can hook into them with Frida.
\figref{}~\ref{fig:novel-decryption-technique} illustrates how this technique is applied on Spatial, an app that enables people to meet through VR~\cite{spatial-app}, as our example.
First (\textbf{Step 1}), we take the app's APK file and extract the version of the Unity framework used to package the app by scanning its configuration files---here we find that the Spatial app uses Unity 2020.1.14f1.
Second (\textbf{Step 2}), we try to find \code{Unity 2020.1.14f1} from a collection of Unity frameworks (we first have to download all versions of Unity onto our system). 
Third (\textbf{Step 3}), we locate \code{mbedtls\_x509\_crt\_verify\_with\_profile()} in the (non-stripped) symbolicated pre-compiled Unity SDK binary file that comes with \code{Unity 2020.1.14f1}.
Subsequently, we extract the binary signature of the certificate validation function, which consists of the 4 bytes preceding the start of the function (\ie \code{\textbf{FD FD FF 17}}) and the first 16 bytes starting from the function address (\ie \code{F7 5B BD A9 F5 53 01 A9} \code{F3 7B 02 A9 F4 03 02 AA}).
We found that we could not use the entire function as our signature due to binary compilation optimizations and stripping.
Fourth (\textbf{Step 4}), we use this binary signature to locate \code{mbedtls\_x509\_crt\_verify\_with\_profile()} in the app's stripped binary file and extract its actual address---for the Spatial app the function is located at address \code{\textbf{0x814468}}.
Finally, we use this extracted address to set a Frida hook for \code{mbedtls\_x509\_crt\_verify\_with\_profile()} in the Frida script (see \figref{}~\ref{fig:traffic-collection}).

\section{Data Types and ATS Details}
 \label{app:ats-data-types-appendix}
 
In~\appref{}~\ref{app:extracting-data-types-appendix}, we provide details about how we identify and group data types, which complements our work in \secref{}~\ref{sec:labeling-data-flow}. In~\appref{}~\ref{app:missed-blocklists}, we provide the full list of potential ATS destinations that are missed by blocklists, which complements our work in \secref{}~\ref{sec:ats-ecosystem}.

\subsection{Extracting Data Types} \label{app:extracting-data-types-appendix}

Please recall that \secref{}~\ref{sec:labeling-data-flow} introduced our methodology for extracting data types from our network traffic dataset. 

Data types can be identified through static values (\eg{} Email, Serial Number, Android ID) which rely on string matching of keywords. On the other hand, dynamic values can change based on the application being tested (\eg{} SDK Version), which rely on a combination of string matching and regular expressions.
\tabref{}~\ref{tab:data-type-extraction-appendix} provides the details on the keywords and regular expressions that we use to extract data types. For instance, to capture different versions of Unity SDK Versions being exposed, we rely on the regular expression \verb"UnityPlayer/[\d.]+\d".

Our 21 data types are groups of other finer grain data types, detailed in~\tabref{}~\ref{tab:data-type-extraction-appendix}. For example, the data type SDK Version considers both Unity and Unreal versions, while Usage Time considers the Start Time and Duration of app usage. Grouping of data types allows us to provide a more complete picture of data collection on \oculusvr{}.


\begin{table*}[htbp]
	\footnotesize
	\centering
	\begin{tabularx}{\linewidth}{ l | p{4cm} | X }
	    \toprule
		\textbf{PII} & \textbf{Finer Grain Data Types} &  \textbf{Keywords or ~\texttt{Regular Expressions}} 
		 \\
		\midrule
        Android ID & - &
          (hard-coded Android ID), android\_id, x--android--id 
        \\ \midrule
        Device ID & - &
          (hard-coded Oculus Device ID), deviceid, device\_id, device--id
        \\ \midrule
        Email & - & (hard-coded user email address), email  \\ \midrule
	    Geolocation & Country Code, Time Zone, GPS & countryCode, timeZoneOffset, gps\_enabled  \\ \midrule
		Person Name & First Name, Last Name & (hard-coded from Facebook Account)  \\ \midrule
		Serial Number & - & (hard-coded Oculus Serial Number), x--oc--selected--headset--serial \\ \midrule
		User ID & User ID and PlayFab ID & user\_id, UserID, x--player, x--playeruid, profileId, anonymousId, PlayFabIDs \\
		\midrule	\textbf{Fingerprint} \\ 
		\midrule
        App Name & App Name and App Version & app\_name, appid, application\_name, applicationId, X--APPID, gameId, package\_name, app\_build, localprojectid, android\_app\_signature, gameVersion, package\_version \\ \midrule
        Build Version & - & build\_guid, build\_tags  \\ \midrule 
        Cookies & - & cookie \\ \midrule
        Flags & Do Not Track, Tracking, Jail Break, Subtitle On, Connection Type, Install Mode, Install Store, Scripting Backend  &  x--do--not--track, tracking, rooted\_or\_jailbroken, rooted\_jailbroken, subtitles, connection--type, install\_mode, install\_store, device\_info\_flags, scripting\_backend \\ \midrule
        Hardware Info & Device Model, Device RAM, Device VRAM, CPU Vendor, CPU Flags, Platform CPU Count and Frequency, GPU Name,  GPU Driver, GPU Information, OpenGL Version, Screen Resolution, Screen DPI, Fullscreen Mode, Screen Orientation, Refresh Rate, Device Info, Platform & 
        device\_model, device\_type, enabled\_vr\_devices, vr\_device\_name, vr\_device\_model,
        Oculus/Quest/hollywood, \verb "Oculus[+ ]?Quest", \verb "Quest[ ]?2", device\_ram, device\_vram, Qualcomm Technologies, Inc KONA, ARM64 FP ASIMD AES, ARMv7 VFPv3 NEON, cpu\_count, cpu\_freq,
        ARM64+FP+ASIMD+AES, arm64-v8a,+armeabi-v7a,+armeabi, Adreno (TM) 650, GIT@09c6a36, GIT@a8017ed,
        gpu\_api, gpu\_caps, gpu\_copy\_texture\_support, gpu\_device\_id, gpu\_vendor\_id, gpu\_driver, gpu\_max\_cubemap\_size, gpu\_max\_texture\_size, gpu\_shader\_caps, gpu\_supported\_render\_target\_count, gpu\_texture\_format\_support, gpu\_vendor, gpu\_version, OpenGL ES 3.2, \verb "\+3664,\+1920" , 3664 x 1920, 3664x1920, width=3664, screen\_size, screen\_dpi, is\_fullscreen, screen\_orientation, refresh\_rate,
        device\_info\_flags, releasePlatform, platform, platformid
        \\ \midrule
        SDK Version & Unity Version, Unreal Version, Client Library, VR Shell Version & 
        \verb "Unity[- ]?v?20[12]\d\.\d+\.\d+", \verb "Unity[- ]?v?[0-6]\.\d+\.\d+", \verb "UnityPlayer/[\d.]+\d", \verb "UnitySDK-[\d.]+\d", x--unity--version, sdk\_ver, engine\_version, X--Unity--Version, sdk\_ver\_full, ARCore,
        X-UnrealEngine-VirtualAgeStats, engine=UE4, UE4 0.0.1
        clientLib, clientLibVersion, x--oc--vrshell--build--name
        \\ \midrule
        Session Info & App Session, Session Counts, Events, Analytics, Play Session Status, Play Session Message, Play Session ID & AppSession, session\_id, sessionid, event-id, event\_id, objective\_id, event-count, event\_count, session-count, session\_count, analytic, joinable, lastSeen, join\_channel, JoinParty, SetPartyActiveGameOrWorldID, SetPartyIDForOculusRoomID, JoinOpenWorld, partyID, worldID, gameOrWorldID, oculusRoomID \\ \midrule
        System Version & - & (hard-coded OS version strings), x-build-version-incremental, os\_version, operatingSystem, os\_family \\ \midrule
        Usage Time & Start Time, Duration & t\_since\_start, startTime, realtimeDuration, seconds\_played, game\_time, gameDuration \\ \midrule
        Language & - & language, language\_region, languageCode, system\_language \\
        
		\midrule	\textbf{VR Sensory Data} \\ 
		\midrule
        VR Field of View & - & vr\_field\_of\_view  \\ \midrule
        VR Movement & Position, Rotation, Sensor Flags & vr\_position, vr\_rotation, gyroscope, accelerometer, magnetometer, proximity, sensor\_flags, left\_handed\_mode  \\ \midrule
        VR Play Area & Play Area, Play Area Geometry, Play Area Dimention, Tracked Area Geometry, Tracked Area Dimension & vr\_play\_area\_geometry, vr\_play\_area\_dimension, playarea, vr\_tracked\_area\_geometry, vr\_tracked\_area\_dimension \\ \midrule
        VR Pupillary Distance & - & vr\_user\_device\_ipd   \\

	    \bottomrule
	\end{tabularx}
	\caption{\textbf{Extracting data types.} Summarizes how we group data types and the keywords and regular expressions (italicized) that we use to identify them (\secref{}~\ref{sec:network-traffic-dataset} and \appref{}~\ref{app:extracting-data-types-appendix}).}
	\label{tab:data-type-extraction-appendix}
\end{table*}


\subsection{Missed by Blocklists}
\label{app:missed-blocklists}
As \oculusvr{} is an emerging platform, there are currently no specialized \blocklist{s} for it. To facilitate the identification of domains that are potential ATS, we target domains that collect  multiple different data types.
As a result, we extend \tabref{}~\ref{tab:missed-fqdns} from \secref{}~\ref{sec:ats-ecosystem} and provide the full details of domains that were missed by \blocklist{s} in \tabref{}~\ref{tab:missed-fqdns-appendix}.

\begin{table*}[tb!]
	\footnotesize
	\centering
	\begin{tabularx}{\linewidth}{p{6cm} p{2.5cm} p{8cm} }
	    \toprule
		\textbf{FQDN} & \textbf{Organization} & \textbf{Data Types} 
		\\
		\midrule
		bdb51.playfabapi.com, 1c31b.playfabapi.com & Microsoft & Android ID, User ID, Device ID, Person Name, Email, Geolocation, Hardware Info, System Version, App Name, Session Info, VR Movement  \\
		\midrule
		sharedprod.braincloudservers.com & bitHeads Inc. & User ID, Geolocation, Hardware Info, System Version, SDK Version, App Name, Session Info, Language  \\
		\midrule
		cloud.liveswitch.io & Frozen Mountain Software & User ID, Device ID, Hardware Info,  System Version, App Name, Language, Cookie \\
		\midrule
		datarouter.ol.epicgames.com & Epic Games & User ID, Device ID, Hardware Info, SDK version, App Name, Session Info  \\
		\midrule
		9e0j15elj5.execute-api.us-west-1.amazonaws.com & Amazon & User ID, Hardware Info, System Version, SDK Version, Usage Time  \\ 
		\midrule
		63fdd.playfabapi.com & Microsoft & Android ID, User ID, Email, SDK Version, App Name \\ 	\midrule
		us1-mm.unet.unity3d.com & Unity & Hardware Info, System Version, SDK Version, Usage Time \\ 	\midrule
		scontent.oculuscdn.com & Facebook & Hardware Info, System Version, SDK Version \\ \midrule
		api.avatarsdk.com & Itseez3d & User ID, Hardware Info, SDK Version \\	\midrule
		52.53.43.176 & Amazon &  Hardware Info, System Version, SDK Version \\	\midrule
		kingspraymusic.s3-ap-southeast-2.amazonaws.com,
		
		s3-ap-southeast-2.amazonaws.com & Amazon &  Hardware Info, System Version, SDK Version \\	\midrule
		pserve.sidequestvr.com & SideQuestVR &  Hardware Info, System Version, Language \\	\midrule
		gsp-auw003-se24.gamesparks.net,
		
		gsp-auw003-se26.gamesparks.net,
		
		gsp-auw003-se30.gamesparks.net,
		
		live-t350859c2j0k.ws.gamesparks.net & GameSparks &  Device ID, Flags \\	\midrule
		yurapp-502de.firebaseapp.com & Alphabet &  Hardware Info, SDK Version \\	
	    \bottomrule
	\end{tabularx}
	\caption{\textbf{Missed by blocklists continued.} We provide third-party FQDNs that are missed by blocklists based on the number data types that are exposed. This is the full details of Table~\ref{tab:missed-fqdns}.}
	\label{tab:missed-fqdns-appendix}
\end{table*}

\mycomment{
\begin{figure}[t!]
    \centering
        \includegraphics[width=\columnwidth]{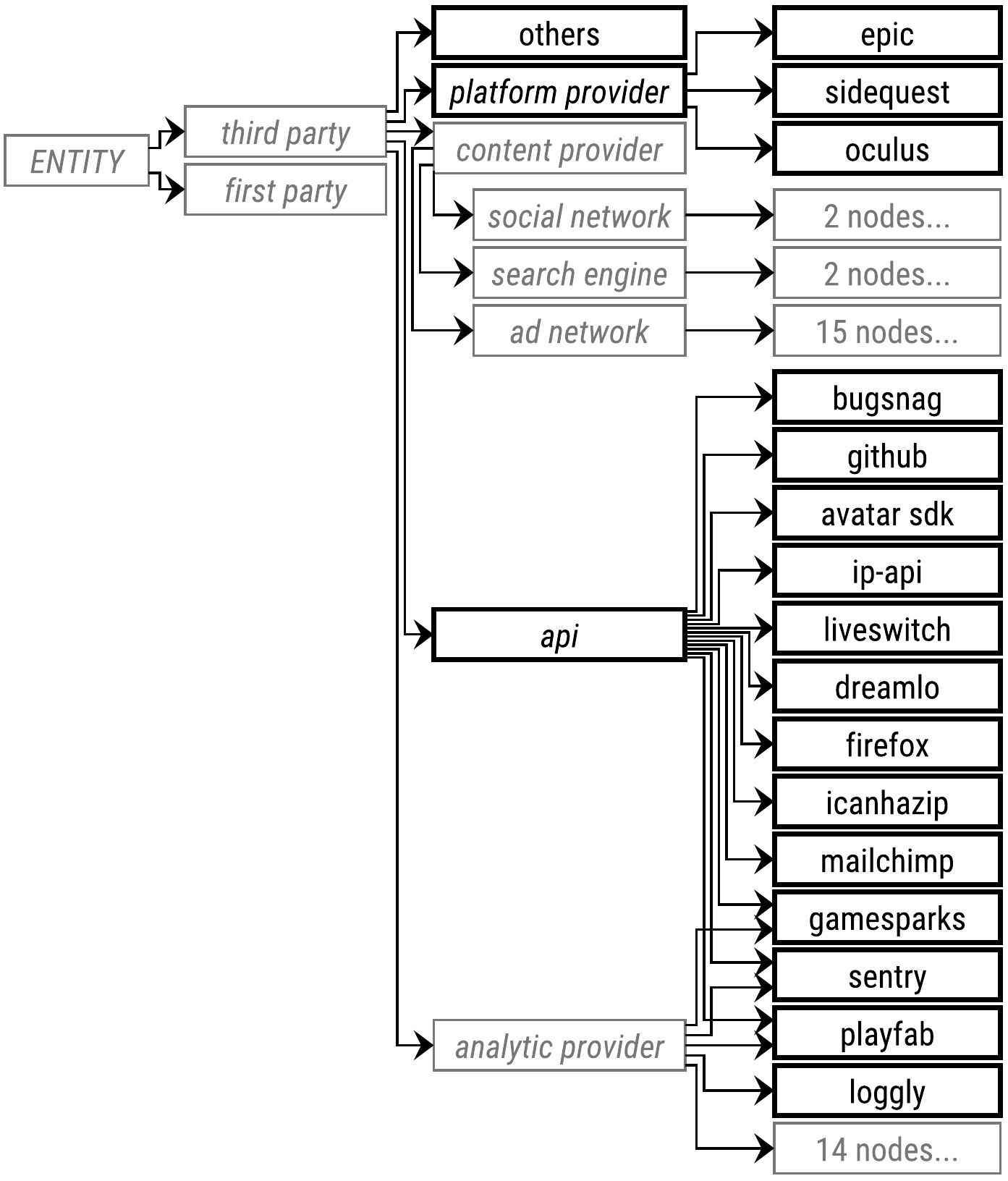}
    \caption{\textbf{Entity Ontology for VR.} Nodes printed in bold are new nodes added on top of the original entity ontology. Nodes with non-italic labels are leaf nodes, which correspond to entity names in data flows.}
    \label{fig:entity-ontology}
\end{figure}}

\section{Privacy Policy Analysis Details}
\label{app:privacy-policy-analysis-details}
In this appendix, we provide more details about \tool{}'s privacy policy analysis we described in \secref{}~\ref{sec:privacy-policy}. 
We describe the details of our improvements for \policheck{} in \appref{}~\ref{app:other-policheck-improvements}, 
our manual validation for \policheck{} in \appref{}~\ref{app:policheck-validation}, and how we integrated \polisys{} into \tool{} (including our manual validation for \polisys{}) in \appref{}~\ref{app:polisis-integration}.

\subsection{Other PoliCheck Improvements}
\label{app:other-policheck-improvements}
In \secref{}~\ref{sec:policheck-polisys}, we mentioned that we have improved \policheck{} in \tool{}. We detail the improvements below.

\myparagraph{Inclusion of third-party privacy policies.} \policheck{} assumes that each app has one privacy policy. In practice, many apps do not disclose third-party data collection clearly in the privacy policies. Instead, they put links to external third-party policies and direct users to read them for more information. For example, consider the following sentence from one of the privacy policies of apps in our dataset: \textit{``For more information on what type of information Unity collects, please visit their Privacy Policy page <link>...''}

\tool{}'s privacy policy analyzer includes statements from external privacy policies if they are referred to in the app's privacy policy. In this case, first-person pronouns (\eg ``we'') in the external privacy policies are translated to the actual entity names (\eg ``Unity'').
Thus, in the above example, the app's data flows are checked against the policy statements extracted both from the app's privacy policy and Unity's privacy policy.

\myparagraph{Resolution of first-party entity names.} Some privacy policies use full company names to refer to the first party, while \policheck{} only considers first-person pronouns (\eg ``we'') as  indications of first-party references. Thus, we found that \policheck{} wrongly recognizes these company names as third parties. As a result, first-party data flows of these apps were wrongly classified as omitted disclosure.

To fix this issue, \tool{} privacy policy analyzer uses a per-app list of first-party names---this list was extracted from: (1) package names, (2) app store metadata, and (3) special sentences in privacy policies such as titles, the first occurrence of a first-party name, and copyright notices. These names are treated as first party.

\myparagraph{Entities.} Entities are names of companies and other organizations. We translate domains to entities in order to associate data flows with disclosures in the privacy policies in Section~\ref{sec:privacy-policy}.

Similar to~\cite{andow2020actions}, we use a manually-crafted list of domain-to-entity mappings to determine which entity that each domain belongs to. For example, \texttt{*.playfabapi.com} is a domain of the entity Playfab. We started from the original \policheck{}'s mapping, and added missing domains and entities to it. We visited each new domain and read the information on the website to determine its entity. If we could not determine the entity for a domain, we labeled its entity as \textit{unknown third party}.
\figref{}~\ref{fig:entity-ontology} displays a partial view of our entity ontology.

\begin{table}[t!]
	\tabletextsize
	\centering

    \minorrevision{
	\begin{tabularx}{\linewidth}{X  r r r r}
	    \toprule
		\textbf{Label} & \textbf{Prec.} & \textbf{Recall} & \textbf{F1} & \textbf{Support}
		\\
		\midrule
		\multicolumn{5}{l}{\textbf{three-class classification}}\\
		\midrule
		consistent     & 0.93 & 0.74 & 0.82 & 454 \\
		incorrect      & 0.50 & 1.00 & 0.67 & 2  \\
		omitted        & 0.77 & 0.94 & 0.85 & 425\vspace{2px}\\
		\textit{macro-average} & 0.74 & 0.89 & 0.81 & \\
		\rowcolor{brandeisblue!10}\textit{micro-average} & 0.84 & 0.84 & 0.84 &  \\
		
		\midrule
		\multicolumn{5}{l}{\textbf{binary classification}} \\
		\midrule
		inconsistent (positive) & 0.77 & 0.94 & 0.85 & 427 \\
		consistent (negative)   & 0.93 & 0.74 & 0.82 & 454 \\
	    \bottomrule
	\end{tabularx}
	}
	\caption{\minorrevision{\textbf{\policheck{} validation.} Multi-class and binary classification metrics for each disclosure type along with the averaged performance. Note that support is in terms of number of data flows.}}
	\label{tab:policheck-eval}
\end{table}


\subsection{PoliCheck Validation}
\label{app:policheck-validation}

We briefly described our manual validation for \policheck{} in \secref{}~\ref{sec:policheck-results}.
\minorrevision{
To test the correctness of \tool{}'s privacy policy analyzer, which is based on \policheck{} that was ported into the VR domain, we followed the methodology described in the \policheck{} paper~\cite{andow2020actions} and another study that applies \policheck{} on Alexa skills~\cite{lentzsch2021heyalexa}. They sampled a portion of consistency results and manually read through the corresponding privacy policies to validate the results.

In \policheck{}, network-to-policy consistency analysis is a single-label five-class classification task. To mitigate biases from human annotators, \policheck{} authors skipped ambiguous disclosures and did not differentiate between clear and vague disclosures during manual validation, which turned it into a three-class (\ie consistent, omitted, and incorrect) classification task. We followed this validation methodology and obtained the complete results that are shown in \tabref{}~\ref{tab:policheck-eval}. The authors reported micro-averaged precision~\cite{andow2020actions}. For completeness and consistency with \policheck{} results, we also report recall, F1-score, and macro-averaged metrics. Micro- and macro-averaging are both popular methods to calculate aggregated precision and recall in multi-class classification tasks \cite{macro-vs-micro}. Macro-averaged precision/recall simply reports the averaged precision/recall of each class. For example, macro-averaged precision is

\begin{equation*}
Pr_{\mathrm{macro}} = \frac{1}{N}(Pr_1 + Pr_2 + ... + Pr_N)
\end{equation*}

\noindent where $N$ is the number of classes and $Pr_{i}$ is the precision of class $i$. In contrast, micro-averaging sums numbers of true positives and false positives of all classes first, and then calculates the metrics. Thus, micro-averaged precision is

\begin{equation*}
Pr_{\mathrm{micro}} = \frac{\mathrm{TP}_1 + \mathrm{TP}_2 + ... + \mathrm{TP}_N}{(\mathrm{TP}_1 + \mathrm{TP}_2 + ... + \mathrm{TP}_N) + (\mathrm{FP}_1 + \mathrm{FP}_2 + ... + \mathrm{FP}_N)}
\end{equation*}

\noindent where $\mathrm{TP}_i$ and $\mathrm{FP}_i$ are numbers of true positive and false positive samples of class $i$. In single-label multi-class classification, every misclassification is a false positive for one class and a false negative for other classes. Thus, the denominators in precision and recall are always equal to the population of samples: micro-averaged precision, recall and F1-score are all the same. Micro-averaging is preferable when the distribution of classes is highly imbalanced, which is the case in our dataset.

In addition, we also report, in \tabref{}~\ref{tab:policheck-eval}, the precision, recall and F1-score of the binary classification case, where we only care about whether the data flows are consistent or not with privacy policy statements. In this case, inconsistent flows are seen as positive samples.
}

\begin{table}[t!]
	\tabletextsize
	\centering

    \minorrevision{
	\begin{tabularx}{\linewidth}{X  r r r r}
	    \toprule
		\textbf{Label} & \textbf{Prec.} & \textbf{Recall} & \textbf{F1} & \textbf{Support}
		\\
		\midrule
		additional service feature     & 0.74 & 0.70 & 0.72 & 20 \\
		advertising                    & 0.94 & 1.00 & 0.97 & 16 \\
		analytics research             & 0.91 & 0.80 & 0.85 & 25 \\
		basic service feature          & 0.82 & 0.45 & 0.58 & 20 \\
		legal requirement              & 0.64 & 1.00 & 0.78 & 9 \\
		marketing                      & 0.92 & 0.75 & 0.83 & 16 \\
		merger acquisition             & 0.78 & 0.88 & 0.82 & 8 \\
		personalization customization  & 0.80 & 0.67 & 0.73 & 6 \\
		service operation and security & 0.82 & 0.64 & 0.72 & 14 \\
		unspecific                     & 0.75 & 0.90 & 0.81 & 49 \\
		\midrule
		\textit{macro-average}        & 0.81 & 0.78 & 0.78 &  \\
		\rowcolor{brandeisblue!10}\textit{micro-average}        & 0.80 & 0.79 & 0.79 &  \\
	    \bottomrule
	\end{tabularx}
	}
	\caption{\minorrevision{\textbf{\polisys{} validation.} Multi-label classification metrics for each purpose along with the averaged performance. Note that support is in terms of number of text segments. Text segments that \polisys{} does not annotate with a purpose is annotated as ``unspecific''.}}
	\label{tab:polisis-eval}
\end{table}

\subsection{Polisis Integration and Validation}
\label{app:polisis-integration}

We described how we used \polisys{} for purpose extraction in \secref{}~\ref{sec:polisis-context}.
\polisys{} is available as a black-box online privacy policy analysis service (\url{https://pribot.org/}). We feed privacy policies (in HTML format) into \polisys{} and get text segments annotated with purposes via \polisys{} Web API. To the best of our knowledge, it internally uses end-to-end deep learning classifiers to annotate purposes at text-segment level~\cite{harkous2018polisis}, which is different from \policheck{}'s sentence-level NLP technique. Since \polisys{} is not open-sourced, we know very little about how \polisys{} segments and processes text internally.

We developed a translation layer to associate \tool{} data flows with purposes from \polisys{}. The translation layer emulates \policheck{}'s text processing on \polisys{} text segments to break them into sentences. Next, it compares binary bag-of-words representation to match sentences from \policheck{} with sentences from \polisys{}. Two sentences from both sides match if one sentence contains all the words in the other sentence. A successful match yields \textit{data type} and \textit{destination} from \policheck{}, and \textit{purpose} from \polisys{}. Although we made the sentence matching very tolerant, it still failed to find some matches due to edge cases caused by the very different text processing pipelines of \policheck{} and \polisys{}.

\minorrevision{\myparagraph{\polisys{} validation.} We evaluated the performance of \polisys{} by manually annotating text segments with purposes. The evaluation process is described in \secref{}~\ref{sec:polisis-context} and the complete results are shown in the upper part of \tabref{}~\ref{tab:polisis-eval}.}

\minorrevision{
\mysection{Responses from Developers}
\label{sec:responses}
We sent courtesy notification emails to inform Oculus and the developers of the 140 apps about our findings on September 13 and 14, 2021. We provide a summary of responses from these developers in~\secref{}~\ref{sec:recommendations}.
Within a period of two weeks, we received 24 responses from these developers: three developers of Oculus free apps, six developers of Oculus paid apps, and 15 developers of SideQuest apps.
Most of these developers (21/24) responded positively and thanked us for sharing our findings about their apps; others responded simply that they have received the message (\eg through an automated reply), or said that the email address we sent our message to was the wrong one.
Five of 19 developers reiterated their position about their data collection practices and/or referred us back to their privacy policy. Notably, 12 of 19 developers inquired further about our findings: they discussed with us to gain deeper insights from our findings, promised to improve their privacy policy, and asked for our advice on how they can write better privacy policies. In particular, some developers expressed the need for training on privacy policy writing and the difficulty in ensuring consistent disclosures---this implicates the need for tools, such as \tool{}.
}

}

\end{document}